\documentclass[11pt]{article}

\usepackage{amstext,amsmath,amssymb,amsfonts,bbm}
\usepackage{hyperref}
\usepackage{authblk}
\usepackage{caption}
\usepackage[latin1]{inputenc}
\usepackage{epsfig}
\usepackage{subfigure}
\usepackage{color}
\usepackage{graphicx}
\usepackage{braket}
\usepackage{slashed}

\usepackage[lmargin=80pt,rmargin=80pt,tmargin=80pt,bmargin=80pt]{geometry}

\newcommand{\be}{\begin{equation}}
\newcommand{\ee}{\end{equation}}
\newcommand{\nn}{\nonumber}
\newcommand{\f}{\frac}
\newcommand{\p}{\partial}

\newcommand{\Tr}{{\rm Tr}}

\newcommand{\la}{\langle}
\newcommand{\ra}{\rangle}
\DeclareMathOperator{\im}{\mathrm{i}}

\let\a=\alpha \let\b=\beta    \let\d=\delta
        \let\l=\lambda
               
\let\s=\sigma \let\t=\tau   \let\vph=\varphi  
    
\let\G=\Gamma     
         
  \let\eps=\epsilon

\newcommand{\bfa}{\mathbf{a}}
\newcommand{\bfb}{\mathbf{b}}
\newcommand{\bfc}{\mathbf{c}}
\newcommand{\bfd}{\mathbf{d}}

\newcommand{\bfm}{\mathbf{m}}
\newcommand{\bfn}{\mathbf{n}}

\newcommand{\tG}{\tilde{G}}

\newcommand{\cA}{\mathcal{A}}

\newcommand{\cF}{\mathcal{F}}
\newcommand{\cG}{\mathcal{G}}

\newcommand{\cK}{\mathcal{K}}

\newcommand{\cS}{\mathcal{S}}

\newcommand{\uG}{\underline{G}}
\newcommand{\uSigma}{\underline{\Sigma}}

\begin{document}

\title{\bf 2PI effective action for the SYK model and tensor field theories}

\author[1]{Dario Benedetti}
\author[2]{Razvan Gurau}

\affil[1]{\normalsize\it Laboratoire de Physique Th\'eorique (UMR 8627), CNRS, Univ.Paris-Sud, \authorcr
\it Universit\'e Paris-Saclay, 91405 Orsay, France \authorcr
email: dario.benedetti@th.u-psud.fr  \authorcr \hfill }

\affil[2]{\normalsize\it CPHT, Ecole Polytechnique, CNRS, Université Paris-Saclay, Route de Saclay, 91128 Palaiseau, France and
Perimeter Institute for Theoretical Physics, 31 Caroline St. N, N2L 2Y5, Waterloo, ON,
Canada \authorcr email: rgurau@cpht.polytechnique.fr \authorcr \hfill}

\date{}

\maketitle

\hrule\bigskip

\begin{abstract}

We discuss the two-particle irreducible (2PI) effective action for the SYK model and for tensor field theories.
For the SYK model the 2PI effective action reproduces the bilocal reformulation 
of the model \emph{without using replicas}.

In general tensor field theories the 2PI formalism is the only way to obtain a bilocal reformulation of the theory, and as such is a precious instrument for the identification of soft modes and for possible holographic interpretations.
We compute the 2PI action for several models, and push it up to fourth order in the $1/N$
expansion for the model proposed by Witten in \cite{Witten:2016iux}, uncovering a one-loop structure in terms of an auxiliary bilocal action.

\end{abstract}

\hrule\bigskip

%\newpage
\tableofcontents

%-------------------------------------------------------------
\section{Introduction}
%-------------------------------------------------------------

The Sachdev-Ye-Kitaev (SYK) model\cite{Sachdev:1992fk,Kitaev}, which is a one-dimensional model of $N$ Majorana fermions with quenched random couplings,
has recently been intensely studied as a testbed for the AdS/CFT duality
\cite{Polchinski:2016xgd,Maldacena:2016hyu,Gross:2016kjj,Kitaev:2017awl}.
Arguably, one of the most useful features of the model is that at the first few orders in $1/N$
it can be reformulated as a bilocal field theory \cite{Kitaev,Sachdev:2015efa,Jevicki:2016bwu}.
This allows to: derive the large-$N$ Schwinger-Dyson equations as equations of motion \cite{Kitaev}; neatly identify the light mode associated to the non-conformal perturbation in the strong coupling limit and derive its effective dynamics (controlled by the so-called Schwarzian action \cite{Maldacena:2016hyu,Jevicki:2016bwu,Kitaev:2017awl}, which can then be matched to a possible gravitational dual \cite{Jensen:2016pah,Maldacena:2016upp,Engelsoy:2016xyb}); efficiently build a perturbative expansion for $n$-point functions of bilinear operators \cite{Jevicki:2016ito,Gross:2017hcz,Gross:2017aos,Kitaev:2017awl}; possibly provide a holographic interpretation along the lines of \cite{Das:2003vw,Koch:2010cy}.
However, the bilocal theory lives in a replica space and it is difficult (and not done in the literature) to study the fluctuations 
that break replica symmetry. Consequently, it only works at the first few orders in $1/N$ at which the quenched and annealed versions of the model coincide 
 \cite{Gurau:2017xhf,Kitaev:2017awl}.

In 2016 Witten \cite{Witten:2016iux} proposed a tensor model \cite{color,RTM} with a similar large-$N$ limit as the SYK model, but without quenched disorder. The fact that this model is a genuine quantum system and has a symmetry which can 
be gauged (thus allowing to restrict the operators in the theory to be singlets) makes it in principle more appealing from the point of view of holography. For these reasons, this model and other similar ones
have been intensely studied at large $N$ \cite{Peng:2016mxj,Klebanov:2016xxf,Klebanov:2017nlk,Giombi:2017dtl,Bulycheva:2017ilt,Beccaria:2017aqc,Choudhury:2017tax,Prakash:2017hwq,Benedetti:2017fmp,Peng:2017kro,Halmagyi:2017leq},
and at finite $N$ \cite{Krishnan:2016bvg,Krishnan:2017txw,Krishnan:2017lra,Krishnan:2018hhu}.
The tensor large-$N$ limit has also been used to derive a new large-$D$ expansion for multi-matrix models
\cite{Ferrari:2017ryl,Azeyanagi:2017drg,Ferrari:2017jgw,Azeyanagi:2017mre}.

While they coincide at leading order, the SYK model and its tensor analogues are quite different at subleading orders
\cite{Bonzom:2017pqs}. Moreover, tensor field theories\footnote{In this paper, tensor field theories are local field theories whose fields transform as a tensor under a global or local symmetry group. They should not be confused with the theories that go under the same name \cite{Rivasseau:2016rgt,BenGeloun:2017jbi} (also sometimes referred to as tensorial group field theories \cite{Geloun:2016qyb}) but which are field theories with a non-local interaction following a tensorial pattern of points identification.} have many more covariant and invariant (or singlet) operators \cite{Geloun:2013kta,BenGeloun:2017vwn,deMelloKoch:2017bvv,Bulycheva:2017ilt}. This has rather drastic consequences: one-dimensional tensor models display a large number of light modes in the infrared
\cite{Bulycheva:2017ilt,Choudhury:2017tax} which are absent in the SYK model. 
In order to study these modes, and possibly to better understand the holographic dual of such theories, it would be useful to have a bilocal reformulation of the theory. In fact, while the construction of the bilocal action in the SYK model is a standard procedure for disordered systems such as spin glasses \cite{Gardner:1985yq,Cugliandolo:2001}, it is not tied to disorder and it can also be understood in a more general context as a special case of the collective field method \cite{Jevicki:1980zg}.
However, until now, no collective field formulation has been found for tensor models (except in the few cases in which an intermediate field representation is possible, 
but which do not have SYK-type behavior \cite{Benedetti:2017fmp}):
in \cite{Choudhury:2017tax} a bilocal action is \emph{postulated} but not derived, while in \cite{Bulycheva:2017ilt} the existence of the new light modes is inferred from other arguments;
in \cite{Dartois:2017xoe} a bilocal action is proposed for the Gurau-Witten model \cite{Witten:2016iux,color}, but it leads to wrong Schwinger-Dyson equations.
Actually, one can expect that no simple and exact reformulation of tensor models is possible in terms of few collective variables because the collective field method of \cite{Jevicki:1980zg} is based on the idea that one could rewrite a theory with a certain symmetry directly in terms of its invariants; but while vector models have only one possible invariant (and its derivatives), and matrix models can be reduced to eigenvalues, which are much less than the original number of variables, tensor models have a much larger number of invariants and no useful reduction to eigenvalues is known.
Furthermore, for vector models the collective field reformulation reduces the large-$N$ expansion to a simple saddle-point (or loop) expansion, which we do not expect to be the case for tensor models.

\

In this paper we propose to use the two-particle irreducible (2PI) effective action formalism \cite{Cornwall:1974vz} (see \cite{Berges:2004yj} for a modern review) for tensor models and show that it provides a useful version of the bilocal reformulation.
The 2PI formalism has been applied to a variety of problems (see for example \cite{Blaizot:2000fc,Berges:2004hn,Berges:2005hc,Reinosa:2009tc} and references therein) and it has been shown to be well suited for a $1/N$ expansion in the case of the $O(N)$ model \cite{Berges:2001fi,Aarts:2002dj}.
Nevertheless, it is not one of the most popular formalisms around, and therefore we will review it in a self-contained way in Sec.~\ref{sec:def-vector2PI}, together with its large-$N$ expansion for the $O(N)$ model in Sec.~\ref{sec:vec-largeN}.
The connection to the collective field formalism is in this case straightforward, as we will explain in Sec.~\ref{sec:vec-auxiliary}.

In order to elucidate the usefulness of the 2PI effective action it is worthwhile to apply it to the SYK model first, as we will do in Sec.~\ref{sec:SYK}. It turns out that the 2PI reformulation 
reproduces exactly all the results of \cite{Maldacena:2016hyu,Jevicki:2016bwu,Kitaev:2017awl}, up to the same order in $1/N$ 
\emph{without using replicas}. This being said, the 2PI reformulation in the SYK model has its own drawbacks:
\begin{itemize}
 \item it requires to know explicitly the graphs contributing to each order in $1/N$. While this is exogenous to the formalism, hence not very aesthetically pleasing, the graph analysis has already been done and 
 we are able to use this information to write the 2PI action up to the same order as the usual replica based bilocal action.
  \item it also fails at higher enough orders in $1/N$. This has nothing to do with replicas, although it happens at the same order at which the replica diagonal ansatz breaks down: it has to do with the lack of commutation between going on shell and taking quenched averages (this will be explained in Sec.~\ref{sec:SYK}).
 \end{itemize}

 The main lesson to be drawn from the 2PI reformulation of the SYK model is that the leading and next-to-leading orders in the $1/N$ expansion of the model are exactly the leading order and 
 first loop correction in a loop expansion of the bilocal theory of \cite{Maldacena:2016hyu,Jevicki:2016bwu,Kitaev:2017awl}. This structure \emph{does not survive} at higher orders: the $1/N$ expansion is a loop expansion in the annealed version of the model,
 but not in the quenched one.
 
The main point of our paper is however that the 2PI formalism becomes much more useful in the tensor case, where the issues with the quenched average are absent, and where we do not yet
have an alternative collective field reformulation. We will apply it to the Gurau-Witten \cite{color,Witten:2016iux} and the Carrozza-Tanasa-Klebanov-Tarnopolsky \cite{Carrozza:2015adg,Klebanov:2016xxf} 
models in Sec.~\ref{sec:tensor2PI}. Among other things we will in this way put on a firmer ground the result of \cite{Choudhury:2017tax} by showing that the bilocal action that they postulated is in fact the 
leading-order 2PI effective action. Furthermore, in the Gurau-Witten model we will be able to expand the action up to fourth order in the $1/N$ expansion, highlighting a similar structure among the three 
subleading terms: they all have the form of a logarithm of a determinant, hence they can be interpreted as Gaussian integrals over bilocal fields. Surprisingly, we will see that all such terms can be interpreted as the one-loop correction of an auxiliary bilocal effective action.

\newpage
%-------------------------------------------------------------
\section{2PI effective action for vector models}
\label{sec:def-vector2PI}
%-------------------------------------------------------------

Let us review the definition and properties of the 2PI effective action.
We consider a theory of real bosonic scalar fields  $\vph_\bfa$, where the index $\bfa$ denotes both a space time point and flavor indices\footnote{For example, vector indices $a$ when the fields form a vector representation of some group. For specific models later in the paper we will make the distinction between space time points and other indices explicit, writing for example $\vph_\bfa=\vph_a(x)$.} with classical action ${\bf S}[\vph]$.
We denote functionals by capital bold letters and, in order to simplify notation, we will sometimes omit the arguments of functionals.
Sums, products, Kronecker deltas and traces include both flavor indices and space time points, and repeated indices are  summed. 
We define:
\be \label{eq:W} 
{\bf W}[j,k] =  \ln\int [d\vph] \; \exp\bigg\{  - {\bf S}[\vph]  +  j_\bfa  \vph_\bfa + \f12  \vph_\bfa k_{\bfa \bfb} \vph_\bfb  \bigg \}\; ,
\ee
which is the generating functional of connected moments of a theory with shifted inverse covariance $ \f{\d^2{\bf S}}{\d\vph_\bfa \d\vph_\bfb }[0] - k_{\bfa \bfb}$.
Observe that ${\bf W}[j,k]$ depends only on the symmetric part of $k_{\bfa \bfb}$ which then is assumed to be symmetric in its indices.
Therefore:
\be
 \f{\d k_{\bfa \bfb}}{\d k_{\bfm \bfn}} = \f12 \cS_{\bfa \bfb ; \bfm \bfn} \;,
\ee
where we have introduced the  projector on symmetric matrices:\footnote{In the case of Grassmann fields (for which we will typically use the letters $\psi$ and $\Psi$ instead of $\vph$ and $\phi$) $k_{\bfa \bfb}$ is antisymmetric, hence the derivative evaluates to the antisymmetric
projector: $\cA_{\bfa \bfb ; \bfm \bfn} = \frac{1}{2} (\delta_{\bfa \bfm} \delta_{\bfb \bfn}-\delta_{\bfa \bfn} \delta_{\bfb \bfm})$.}
\begin{equation}  \label{eq:symm-proj}
\cS_{\bfa \bfb ; \bfm \bfn} = \frac{1}{2} (\delta_{\bfa \bfm} \delta_{\bfb \bfn}+\delta_{\bfa \bfn} \delta_{\bfb \bfm})\;.
\end{equation}

To simplify notation we will denote sometimes the functional derivatives as indices: 
\[
{\bf W}_{  j_\bfa } [j,k] \equiv \frac{\d {\bf W}} {\d {j_\bfa} } [j,k], \qquad {\bf W}_{ k_{\bfa \bfb}} [j,k]\equiv \frac{\d {\bf W}} {\d {k_{\bfa \bfb}}} [j,k] \;.
\]
We denote ${\bf \Phi}$ and ${\bf G}$ the connected 1-point and 2-point functions of the theory with sources $j$ and $k$:
\begin{equation}\label{eq:W-eom}
 {\bf \Phi}_\bfa [j,k]  = {\bf W}_{ j_\bfa}[j,k]   \;,
\end{equation}
\be \label{eq:Wjj-Wk}
 {\bf G}_{\bfa \bfb} [j,k]= {\bf W}_{ j_\bfa  j_\bfb}[j,k] = 2 {\bf W}_{  k_{\bfa \bfb}}[j,k] -  {\bf W}_{  j_\bfa}[j,k]{\bf W}_{  j_\bfb}[j,k] \; .
\ee
We are generally interested in the connected 1-point and 2-point functions of the theory without sources, for which we introduce the following notation:
\begin{equation}
 {\bf \Phi}_\bfa[0,0]    = \la \vph_\bfa  \ra_{\rm conn} \equiv \underline{\phi}_\bfa \;, \qquad  {\bf G}_{\bfa \bfb} [0,0]    =   \la \vph_\bfa \vph_\bfb \ra_{\rm conn} \equiv \underline{G}_{\bfa \bfb}\;.  
\end{equation}
Notice that ${\bf G}_{\bfa \bfb}$ (hence in particular $\uG_{\bfa \bfb}$) is symmetric in its indices.

For a  free theory with covariance $C$ we obtain:
\be
 {\bf W}^C[j,k]  =  - \frac{1}{2} \Tr[\ln(C^{-1} - k ) ] + \frac{1}{2} j_\bfa \left( \frac{1}{C^{-1} - k} \right)_{\bfa \bfb} j_\bfb \;.
\ee
and as a consequence:
\be \label{eq:freeGk}
  {\bf \Phi}^C_\bfa [j,k]  = \left( \frac{1}{C^{-1} - k} \right)_{\bfa \bfb} j_\bfb    \;,\qquad  {\bf G}^C_{\bfa \bfb} [j,k] = \left( \frac{1}{C^{-1} - k} \right)_{\bfb\bfa} \; .
\ee

Let $\{ {\bf J}_\bfa[\phi, G] , {\bf K}_{\bfa \bfb}[\phi, G] \} $ be the inverse of $ \{  {\bf \Phi}_\bfa [j,k],{\bf G}_{\bfa \bfb} [j,k] \} $. For a free theory they are:
\[
   {\bf J}^C_\bfa[\phi, G] =  ( G^{-1})_{\bfa \bfb} \phi_\bfb  \qquad   {\bf K}^C_{\bfa \bfb}[\phi, G] = (C^{-1})_{\bfa \bfb} - ( G^{-1})_{\bfa \bfb} \; .
\]
The 1- and 2-point functions $\underline{\phi}$ and $\underline{G} $ are then determined implicitly 
by the equations:
\be
 {\bf J}_\bfa[ \underline{\phi}, \underline{G} ] =0, \qquad {\bf K}_{\bfa \bfb}[ \underline{\phi}, \underline{G} ] =0 \;,
\ee
and for the free theory we get $\underline{\phi}^C = 0,  \underline{G}^C = C $.

The second derivative of ${\bf W}$ with respect to $k$ is:
\begin{align}
 {\bf W}_{  k_{\bfa \bfb} k_{\bfc \bfd} } = \frac{1}{2} {\bf G}_{\bfa \bfb;  k_{\bfc \bfd} } + \frac{1}{2} {\bf \Phi}_{\bfa;k_{\bfc \bfd}}  {\bf \Phi}_{\bfb} +  \frac{1}{2} {\bf \Phi}_{\bfa} {\bf \Phi}_{\bfb;k_{\bfc \bfd}} \;.
\end{align}
Assuming that we are in a symmetric phase in which the 1-point function is zero, $\underline{\phi}=0$, we obtain:
\begin{equation}\label{eq:smecher}
 \sum_{\bfc \bfd} {\bf W}_{k_{\bfa \bfb}k_{\bfc \bfd}} [j=0,k=0] \; {\bf K}_{\bfc \bfd;G_{\bfm \bfn}} [\underline{\phi} =0 , \underline{G}   ] 
 = \f12 \f{\d G_{\bfa \bfb}}{\d G_{\bfm \bfn}} = \f12 S_{\bfa \bfb ; \bfm \bfn} \;.
\end{equation}

The connected 4-point function is the fourth derivative ${\bf W}_{j_\bfa j_\bfb j_\bfc j_\bfd} [0,0] $. It can be re expressed using derivatives with respect to $k$, as 
several relations exist between derivatives of ${\bf W}[j,k]$ with respect to $j$ and $k$. The simplest one which is obtained by
noticing that deriving the partition function $\exp\{ {\bf W}[j,k]\}$ once with respect to $k$ we obtain (one half times) the same result as deriving twice with respect to $j$:
\be
({\bf W}_{ j_\bfa  j_\bfb}+  {\bf W}_{  j_\bfa}{\bf W}_{  j_\bfb} - 2 {\bf W}_{  k_{\bfa \bfb}}) e^{\bf W} =0\;,
\ee
leading to Eq.~\eqref{eq:Wjj-Wk}. Deriving this equality either one more time with respect to $k$ or two more times with respect to $j$, and 
combining the results we obtain a long relation,
which simplifies considerably in a symmetric phase $\underline{\phi}=0$:
\be \label{eq:Wkk}
{\bf W}_{k_{\bfa \bfb} k_{\bfc \bfd}}[0,0] = \f14 \bigg{(} {\bf W}_{j_\bfa j_\bfb j_\bfc j_\bfd} + {\bf W}_{j_\bfa j_\bfc} {\bf W}_{j_\bfb j_\bfd} + {\bf W}_{j_\bfa j_\bfd} {\bf W}_{j_\bfb j_\bfc} \bigg{)}_{j,k=0}
\equiv \f14 \cF_{ (\bfa , \bfb ) ; ( \bfc , \bfd )} \;.
\ee
The function $ \cF_{ (\bfa , \bfb ) ; ( \bfc , \bfd )} $ is the full 4-point function minus the contribution of the disconnected channel $(\bfa, \bfb)(\bfc , \bfd)$.
For example, in the free theory we obtain from \eqref{eq:Wjj-Wk} and \eqref{eq:freeGk}:
\be
{\bf W}^C_{k_{\bfa \bfb} k_{\bfc \bfd}}[0,0] = \f12  C_{\bfb \bfm}  C_{\bfm \bfa} S_{\bfm \bfn ; \bfc \bfd} = \f14 (C_{\bfb \bfc}  C_{\bfd \bfa} + C_{\bfb \bfd}  C_{\bfc \bfa}) \;.
\ee

\

We define the \emph{2PI effective action}\footnote{The reason for this name will become clear below.} of the theory as the double Legendre transform of ${\bf W}[j,k]$:
\be \label{eq:Gamma}
\begin{split}
{\bf \G}[\phi,G] & =  - {\bf W}[{\bf J} , {\bf K} ] +  {\bf J}_\bfa \phi_\bfa + \f12  \phi_\bfa {\bf K}_{\bfa \bfb} \phi_\bfb + \f12 \Tr[G {\bf K} ] 
\; . 
\end{split}
\ee
Deriving \eqref{eq:Gamma} with respect to $\phi$ and $G$, we obtain the two identities:
\be \label{eq:Gamma-deriv}
 {\bf \G}_{\phi_\bfa} [\phi,G] =  {\bf J}_\bfa[\phi,G] + {\bf K}_{\bfa \bfb}[\phi,G] \phi_\bfb  \;, \qquad  {\bf \G}_{G_{\bfa \bfb}} [\phi,G] = \frac{1}{2}  {\bf K}_{\bfb \bfa} [\phi,G] \;.
\ee
Furthermore, ${\bf \G}_{GG} [\phi,G] =  \frac{1}{2}  {\bf K}_G [\phi,G] $ which, combined with  Eq.~\eqref{eq:smecher} and \eqref{eq:Wkk}, yields for a theory in the symmetric phase:
\begin{equation} \label{eq:truc}
 \cF_{ (\bfa , \bfb ) ; ( \bfc , \bfd )} {\bf \G}_{G_{\bfc  \bfd}G_{\bfm \bfn}} [0, \underline{G}]  = \cS_{\bfa \bfb ; \bfm \bfn} \;.
\end{equation}

As usual we get back to ${\bf W}[j,k]$ by means of a new Legendre transform:
\be \label{eq:Gamma-inverse}
\begin{split}
{\bf W}[j,k] & = - {\bf \G}[{\bf \Phi}, {\bf G} ] + j_\bfa {\bf \Phi}_\bfa + \f12  {\bf \Phi}_\bfa k_{\bfa \bfb} {\bf \Phi}_\bfb + \f12 \Tr[ {\bf G} k ]  \;,
\end{split}
\ee
where the functionals ${\bf \Phi}[j,k], {\bf G}[j,k]$ are determined by solving:
\be\label{eq:Gamma-eom-source} 
 {\bf \G}_{\phi}[{\bf \Phi}, {\bf G} ] =  j  + k {\bf \Phi} \;, \qquad  {\bf \G}_{G} [{\bf \Phi}, {\bf G} ] = \frac{1}{2} k \;.
\ee

The 2PI effective action has a number of interesting features \cite{Cornwall:1974vz,Berges:2004yj}:
\begin{enumerate}
\item 
The solution of the equations of motion ${\bf \G}_{\phi} =  0 ,{\bf \G}_{G} = 0$ 
is $\underline{\phi}, \underline{G}$, which are the connected 1- and 2-point functions of the theory.
\item It can be evaluated in a loop expansion.
Substituting $ {\bf J} [ \phi, G]$ and $ {\bf K}[\phi,G]$ for $j$ and $k$ into \eqref{eq:W}, and using Eq.~\eqref{eq:Gamma}, we obtain:
\be
 e^{- \Gamma[\phi, G] +{\bf J} \phi   + \frac{1}{2} \phi {\bf K} \phi  +  \frac{1}{2}\Tr \left[G  {\bf K} \right] 
 } 
 = \int [d\vph] \; e^{    - {\bf S}[\vph]  +  {\bf J} \vph  +  \frac{1}{2} \vph {\bf K} \vph     } \;.
\ee
We translate $\varphi \to \phi+ \varphi$ and expand
\[{\bf S} [\phi + \varphi ] = {\bf S} [\phi] + {\bf S}_{\phi} [\phi] \varphi + \frac{1}{2} \varphi {\bf S}_{\phi\phi} [\phi] \varphi  + {\bf S}_{\rm int}[\phi,\varphi] \;, \]
where the interacting part of the action ${\bf S}_{\rm int}[\phi,\vph]$ contains all higher powers of $\vph$.
We obtain:
 \be\label{eq:start}
   e^{- {\bf \G} [\phi, G]   + \frac{1}{2} \Tr\left[G  {\bf K} \right] 
 } 
 = \int [d\vph] \; e^{   
   - {\bf S} [\phi]  +\left( {\bf J} + {\bf K} \phi - {\bf S}_{\phi}[\phi]   \right)\vph    -\frac{1}{2} \vph \left(   {\bf S}_{\phi\phi}[\phi] - {\bf K}  \right) \vph -  {\bf S}_{\rm int}[\phi,\varphi] } \;,
 \ee
or, using \eqref{eq:Gamma-deriv}:
 \be\label{eq:Gamma-self-consist}
   e^{- {\bf \G} [\phi, G]   + \Tr\big[G {\bf \G}_G [\phi,G] \big] 
 } 
 = \int [d\vph] \; e^{   
   - {\bf S} [\phi]  +\left( {\bf \G}_{\phi} [\phi,G] - {\bf S}_{\phi}[\phi]   \right)\vph    -\vph \left(  \frac{1}{2}  {\bf S}_{\phi\phi}[\phi] - {\bf \G}_G [\phi,G]   \right) \vph -  {\bf S}_{\rm int}[\phi,\varphi] } \;.
 \ee
So far Eq.~\eqref{eq:start} is exact. In order to evaluate it at one loop, we observe that at the classical level $ {\bf \G} [\phi, G]  \approx  {\bf S} [\phi]$, hence 
$    {\bf J} + {\bf K} \phi - {\bf S}_{\phi}[\phi]  = {\bf \G}_{\phi }  - {\bf S}_{ \phi }  $ is already at one loop
and the linear term in the Gaussian integral can be neglected as it yields a two loop effect upon integration over $\varphi$. Thus at one loop we have:
\be\label{eq:gamaoneloop}
 {\bf \G}_1 [\phi, G]   =  {\bf S} [\phi]  + \frac{1}{2} \Tr\left[G  {\bf K}_1 \right]   +  \frac{1}{2}\Tr\ln \big({\bf S}_{\phi\phi}[\phi] - {\bf K}_1  \big) \;,
 \ee
 where ${\bf K}_1$ is the functional ${\bf K}[\phi, G] = 2 {\bf \G}_G [\phi,G]  $ evaluated at one loop. On the other hand, at one loop:
\be 
 \frac{\d {\bf  \G}_1 [\phi, G]}{\d G} -  \frac{1}{2} {\bf K}_1 = \frac{1}{2}\left(  G - \frac{1}{ {\bf  S}_{\phi\phi}[\phi] - {\bf K}_1 } \right) \frac{\d {\bf K}_1}{ \d G } = 0  \;,
\ee
which in turn fixes ${\bf K}_1 =  {\bf  S}_{\phi \phi}[\phi] - G^{-1} $. Substituting this in Eq.~\eqref{eq:gamaoneloop} and discarding a constant term we 
obtain:\footnote{Repeating the same constructions for complex or Grassmann fields, it is easy to see that we arrive at a similar expression, but with the functional trace terms multiplied 
by an extra factor 2 for the complex case and by a minus for the Grassmann case.}
\be \label{eq:1loop+rest}
{\bf \G}[\phi,G] = {\bf S}[\phi]  +\f12 \Tr[\ln G^{-1}] + \f12 \Tr[G_0^{-1}G] + {\bf \G}_2[\phi,G] \;,
\ee
where $G_0 = ( {\bf S}_{\phi\phi}[\phi] )^{-1}$ is the free covariance of the theory around the field configuration $\phi$ and ${\bf \G}_2[\phi,G]$ starts at two loops. 

In the free theory of covariance $C$, the one loop result is exact, and therefore we have:
\be
{\bf \G}^C[\phi,G] = \f12 \phi_\bfa C^{-1}_{\bfa \bfb} \phi_\bfb +\f12 \Tr[\ln G^{-1}] + \f12 \Tr[C^{-1}G]  \;,
\ee
and it can be easily verified that \eqref{eq:truc} holds.

\item The equations of motion of \eqref{eq:1loop+rest} with respect to $G$ write: 
\be
 G^{-1} = G_0^{-1} +2 \frac{\d {\bf \G}_2}{ \d G }  \; .
\ee
As $G_0$ is the free covariance of the theory and $G$ is the connected two point function, it follows 
form the standard Schwinger-Dyson equation $  G^{-1} = G_0^{-1} - \Sigma $
that $ -2 \frac{\d {\bf \G}_2}{ \d G } $ must be identified with the self energy $\Sigma$ of the model,
which is the sum of amputated one-particle-irreducible two point graphs.
\item ${\bf \G}_2[\phi,G]$ is given by (minus) the sum of all the two-particle irreducible vacuum graphs (i.e. graphs that do not disconnect when cutting open any two edges) with vertices given by the effective interaction
$ {\bf S}_{\rm int}[\phi,\varphi]$ and effective propagators $G$. 
This is slightly non trivial. From Eq.~\eqref{eq:start} we see that ${\bf \G}[\phi,G] $ is the sum of connected graphs with:
\begin{itemize}
 \item trivalent or higher order vertices given by $ - {\bf S}_{\rm int}[\phi,\varphi]$,
 \item univalent vertices $   \varphi( {\bf J} + {\bf K} \phi - {\bf S}_{\phi}[\phi])$,
 \item propagators $ \left(   {\bf S}_{\phi\phi}[\phi] - {\bf K}  \right)^{-1}$,
 \item a vacuum term $ \frac{1}{2} \Tr\left[G  {\bf K} \right]$.
\end{itemize}
 
On the other hand, as $ -2 \frac{\d {\bf \G}_2}{ \d G } = \Sigma $ it follows that ${\bf \G}_2$ can be reconstructed by reconnecting the two external vertices of the self energy by a 
propagator $G$, and since $\Sigma$ is one-particle-irreducible (1PI), ${\bf \G}_2$ is two-particle-irreducible (2PI).

Two questions arise:
\begin{itemize}
 \item what happened to the  univalent vertices? As ${\bf \G}_2$ is 2PI it is in particular 1PI, hence can not have any univalent
 vertices. What happens is that the perturbative expansion of ${\bf \G}[\phi,G]$ in Eq.~\eqref{eq:start} for generic ${\bf J}$ and ${\bf K}$ is built out of connected graphs. Each connected graph has the structure of a tree connecting 1PI 
 vertex kernels. At the self-consistent values of the sources ${\bf J}$ and ${\bf K}$, obtained from \eqref{eq:Gamma-deriv}, the univalent vertices $ \varphi( {\bf J} + {\bf K} \phi )  $
 act as counter terms and subtract the contribution of all the trees with more that one 1PI vertex kernel. 
 
 \item the covariance of the theory is  $ \left(   {\bf S}_{\phi \phi}[\phi] - {\bf K}  \right)^{-1}$, so why are the edges of the 2PI graphs contributing to ${\bf \G}_2$ decorated by $G$?
 The 1PI kernels can still be two-particle reducible. However, one can resum all the two-point function corrections and replace the propagators by the full two point function 
 of the  theory which is $G$. Using the resummed two point function makes the graphs 2PI. 
\end{itemize}
 
\end{enumerate}

In summary we can write the schematic expression:
\be \label{eq:2PI-pathintegral}
   e^{- {\bf \G} [\phi, G]   } 
 = e^{   - {\bf S} [\phi] - \frac{1}{2} \Tr\left[G_0^{-1} G\right]} \int_{2PI} [d\vph] \; e^{   
     -\frac{1}{2} \vph G^{-1}\vph -  {\bf S}_{\rm int}[\phi,\varphi] } \;,
 \ee
where the subscript 2PI reminds us that in the perturbative expansion of the functional integral we only retain 2PI graphs.

%-------------------------------------------------------------
\subsection{Large-$N$ expansion}
\label{sec:vec-largeN}
%-------------------------------------------------------------

While the properties listed above are completely generic, we are now going to review a useful expansion  of the 2PI effective action that is applicable to certain models, namely the $1/N$ expansion. 
We will use a classical example \cite{Cornwall:1974vz,Berges:2004yj}, the $O(N)$ model, in which $N$ is the number of scalar fields:  $\vph_a(x)$, with $a=1,\ldots,N$.
From now on, we make explicit the vector indices and the space time points. We write $\int_x = \int d^d x$,  $\int_{x,y} = \int d^dx  d^dy$, and so on and 
we denote by $\Tr$ a trace both on vector indices and a functional trace, i.e. for a matrix-valued bi-local field $F_{ab}(x,y)$ we have 
\be\nn
\Tr[F] = \int_{x,y}\, \d(x-y)\, \Tr[F_{ab}(x,y)] = \int_x\, F_{aa}(x,x)\;.
\ee
As before, summation is implicit on repeated indices. 

In the $O(N)$ model, the $N$ scalars are postulated to transform in the fundamental representation of the $O(N)$ group, i.e.
\be
\vph_a(x) \to R_{ab} \vph_b(x) \;, \;\;\;\;\;  R\in O(N)\;,
\ee
and the action is chosen to be invariant under such transformations. More specifically, restricting to quartic interactions, the action is:
\be \label{eq:vec-mod}
{\bf S} [\vph]=  \f12  \int_{x,y}  \vph_a(x) C^{-1}(x,y) \vph_a(y)   +\f{\l}{4! N} \int_x (\vph_a(x)\vph_a(x))^2 \;,
\ee
where $C(x,y)$ is the covariance of the Gaussian functional measure of the free theory. 
In $d\geq 1$, $C^{-1}(x,y)$ is usually the kernel of a differential operator, e.g. $C^{-1}= -\p^2+m^2$.

All the definitions we introduced above for the 2PI effective action apply directly, with:
\be \label{eq:vec-G0}
G_{0,ab}^{-1}(x,y) = C^{-1}(x,y)\d_{ab}+ \f{\l}{6 N} (\phi_c\phi_c) \d_{ab}\d(x-y) + \f{\l}{3 N} \phi_a\phi_b \d(x-y) \;,
\ee
\be \label{eq:vec-int}
{\bf S}_{\rm int}[\phi,\vph] = \int_x \left(\f{\l}{6 N} \phi_a\vph_a \vph_b\vph_b+ \f{\l}{4! N} (\vph_a\vph_a)^2 \right)\;.
\ee

\begin{figure}[htb]
 \begin{center}
 \includegraphics[width=8cm]{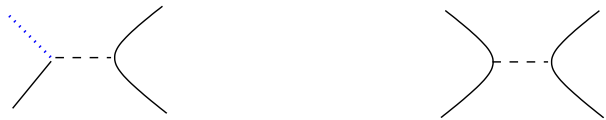}  
 \caption{The two vertices from Eq.~\eqref{eq:vec-int}.} \label{fig:vertices}
 \end{center}
 \end{figure}

In order to construct the $1/N$ expansion, one should take into account the implicit $N$-dependence due to the presence of $N$ variables.
This is done by counting any ``single-trace'' invariant as contributing with a factor $N$. There are two types of such invariants in the $O(N)$ model: $\Tr[G^n]$ and $\phi_a(G^n)_{ab}\phi_b$.
Taking into account also the explicit factor $N^{-1}$ in the coupling, one immediately finds that the first three terms in:
\be
{\bf \G}[\phi,G] = {\bf S}[\phi]  +\f12 \Tr[\ln G^{-1}] + \f12 \Tr[G_0^{-1}G] + {\bf \G}_2[\phi,G]  \;,
\ee
all scale like $N$, except the contribution from the last term in \eqref{eq:vec-G0} which is of order one. 
Next, one observes that the last term can be expanded as:
\be \label{eq:vec-expansion}
 {\bf \G}_2[\phi,G] =  {\bf \G}^{ ( 1 ) }_2[\phi,G]  + {\bf \G}^{( 0 ) }_2[\phi,G] + {\bf \G}^{ (-1) }_2[\phi,G] +\dots \;,\qquad \text{with}\;\; {\bf \G}^{ (p) }_2[\phi,G] \sim N^p \;.
\ee
This is again somewhat non trivial. We first review the Feynman expansion of $ {\bf \G}_2[\phi,G]$ for the $O(N)$ model. In \eqref{eq:vec-int} there are two kinds of vertices, a trivalent and
a tetravalent one, which we represent in Fig.~\ref{fig:vertices}. The solid lines track the identification of the indices in the vertex. The dashed edge symbolizes the vertex (in an intermediate 
field representation it would correspond to the propagator of the intermediate field), and the blue dotted halfedge represents the background field $\phi$.

\begin{figure}[htb]
 \begin{center}
 \includegraphics[width=3cm]{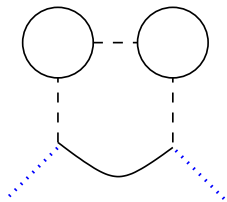}  
 \caption{An example of a graph with one tetravalent and two trivalent vertices.} \label{fig:graph}
 \end{center}
 \end{figure}

 The vertices are connected by propagators $G$ which connect the solid half edges into solid edges. An example of a Feynman graph is presented in Fig.~\ref{fig:graph}.
 
 We notice first of all that ${\bf \G}_2[\phi,G]$, and hence ${\bf \G}[\phi,G]$, will only contain even powers of $\phi$. This is obvious at one-loop level, because there are only even powers 
 in the action \eqref{eq:vec-mod}, and of course in the second variation \eqref{eq:vec-G0}. And is also true beyond one loop, because a single $\phi$ is attached to each 3-valent vertex, while
 4-valent vertices carry no powers of $\phi$. Therefore, any closed graph will necessarily contain an even number of $\phi$ fields.
As a consequence, the equation of motion $\d{\bf \G}/\d\phi=0$ admits the solution $\phi=0$, which is the only solution giving an invariant 1-point function. Whether such a solution is stable or not, 
and whether there are other stable solutions, will depend on the space dimension. In particular, in $d\leq 2$ spontaneous symmetry breaking of continuous symmetries is impossible \cite{Mermin:1966fe,Coleman:1973ci},
hence we do not expect other stable solutions for $\phi$.\footnote{However, one should keep in mind that the large-$N$ limit can sometimes lead to an apparently opposite conclusion, as explained for the chiral Gross-Neveu model by Witten \cite{Witten:1978qu}. See also \cite{Benedetti:2017fmp} for an analogue phenomenon in a tensor-valued version of the Gross-Neveu model.}
 
 Coming back to the $1/N$ expansion, from the Feynman rules one obtains a trace over the vector indices of $G$ to some power for each closed loop of the solid strands, hence each such loop should be counted as 
 a factor $N$ (as seen for example by taking $G_{ab}\sim \d_{ab}$).
 Each vertex brings instead a factor $1/N$.  The open strands connect pairwise the background fields $\phi$, and correspond to scalar product of the type $\phi_a(G^n)_{ab}\phi_b$, so they also should be counted 
 as a factor $N$ (as seen for example by taking $\phi_a\phi_b\sim G_{ab}\sim  \d_{ab}$).

 The power counting in $N$ is transparent in a \emph{loop vertex representation} \cite{Magnen:LVE} (or \emph{cacti} representation) in intermediate field. 
 The loops of vector indices are contracted into \emph{loop vertices} (of arbitrary degree) and the original Feynman vertices become edges of the intermediate field (the black dashed edges in Fig.~\ref{fig:graph}).
 The open strands can be contracted to \emph{external} vertices (of degree two). 
 In this representation the scaling with $N$ of a graph is $N^{- E + L+ L_{\rm ext}}$ where $E$ is the number of intermediate field edges (i.e. vertices in the original Feynman representation), $L$ the number 
 of loop vertices, and $ L_{\rm ext}$ the number of external vertices (i.e.  half the number of background fields). As the graph is connected, the number of excess edges\footnote{Loops (in the physics literature) for the intermediate field.} 
 in the intermediate field is $E - (L+ L_{\rm ext} )+1 =  \omega \ge 0$.
 
\begin{figure}[htb]
 \begin{center}
 \includegraphics[width=3cm]{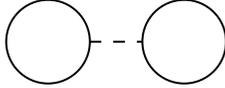}  
 \caption{The leading order contribution of the large-$N$ expansion in the $O(N)$ model.} \label{fig:vectorLO}
 \end{center}
 \end{figure}

 It follows that the scaling in $N$ of a graph is $N^{ -\omega  + 1 }$, hence the graphs contributing to ${\bf \G}_2$ scale at most like $N$ and they scale like $N$
 only if they are trees in the intermediate field. Furthermore, the graphs contributing to  ${\bf \G}_2[\phi,G]$ must at the same time be 2PI  from the point of view of the original propagators, which translates into 
 the constraint that the tree has no vertices of degree greater than one.
 Thus only one graph (the double tadpole of Fig.~\ref{fig:vectorLO}) contributes at leading order (LO):
\be
 {\bf \G}^{(1)}_2[\phi,G] = \f{\l}{4! N}  \int_x G_{aa}(x,x) G_{bb}(x,x) = N \f{\l}{4!}  \int_x G(x,x)^2 \; ,
\ee
which in particular is independent of $\phi$. In the last expression in order to make the $N$-dependence more explicit we restricted to $G_{ab}(x,y) = G(x,y) \d_{ab}$, which is valid on shell in the symmetric phase. 

The equations of motion of ${\bf \G}[\phi,G]$ at LO read:
\be \label{eq:vec-eom1}
0= (C^{-1}\phi_a)(x) + \f{\l}{6N} \phi_a(x) (\phi_b(x)\phi_b(x)) + \f{\l}{6 N} \phi_a(x) G_{bb}(x,x) %+ \f{\l}{3 N} G_{ab}(x,x)\phi_b(x) 
\;,
\ee
\be \label{eq:vec-eom2}
\begin{split}
 G^{-1}_{ab}(x,y) 
   = \left(C^{-1}(x,y)+  \f{\l}{6 N} G_{cc}(x,x)  \delta(x-y)+ \f{\l}{6 N} (\phi_c\phi_c) \d(x-y) \right)\delta_{ab}
 %+ \f{\l}{3 N} \phi_a\phi_b \d(x-y) 
 \;.
\end{split}
\ee 
The first one clearly admits the $\phi_a=0$ solution, which plugged back into the second equation leads to the large-$N$ Schwinger-Dyson (SD) equations for the 2-point 
function:\footnote{We notice that in $d=0$ (and fixing for example $C=1$) the SD equation becomes a simple quadratic equation for $G$ with solution $G=3(-1\pm \sqrt{1+2\l/3})/\l$, thus exhibiting a well-known singularity 
at a negative value of the coupling (e.g. \cite{DiVecchia:1990ce}). For $d\geq 1$ instead (with $C^{-1}=-\p^2+m^2$), the SD equation simply leads to a renormalization of the mass (a finite one in $d=1$).}
\be
G^{-1}_{ab}(x,y) =  \left(C^{-1}(x,y) +  \f{\l}{6 N} G_{cc}(x,x) \delta(x-y)\right) \delta_{ab}\;.
\ee
As anticipated, we find that $G_{ab}=G(x,y)\d_{ab}$ on shell, with the scalar part satisfying $G^{-1}(x,y) = C^{-1}(x,y) +  \f{\l}{6} G(x,x) \delta(x-y)$. Notice 
also that a nonzero solution for $\phi_a$ (necessarily a constant solution because of translation invariance) implies that it is a zero mode of the inverse 2-point 
function, i.e. $(G^{-1}_{ab}\phi_b)(x)=0$: this is because in the case of spontaneous symmetry breaking we have $N-1$ Goldstone modes and only one radial mode, but by 
keeping only LO terms we have discarded the latter. 
Including subleading terms it is no longer
true that the equation of motion of $\phi$ can be written as $(G^{-1}_{ab}\phi_b)(x)=0$: in particular in the LO approximation 
we have discarded a $\f{\l}{3 N} G_{ab}(x,x)\phi_b$ term in \eqref{eq:vec-eom1} and a $\f{\l}{3 N} \phi_a\phi_b \d(x-y) $ term in \eqref{eq:vec-eom2}, both coming from the one-loop part of the 2PI effective action.

\begin{figure}[htb]
 \begin{center}
 \includegraphics[width=12cm]{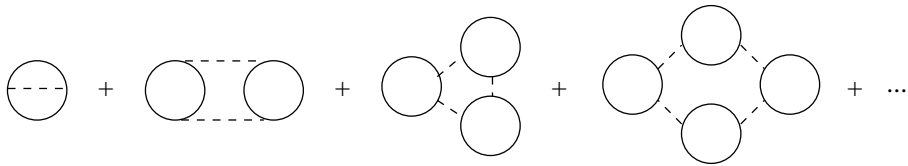}  
 \caption{The next-to-leading order contribution of the large-$N$ expansion in the  $O(N)$ model.} \label{fig:vectorNLO}
 \end{center}
 \end{figure}

At next-to-leading order (NLO), and for $\phi=0$, we have graphs with $E=L$, always with the 2PI restriction: they are the closed chains of bubbles depicted in Fig.~\ref{fig:vectorNLO}. 
The form an infinite family, but thanks to their simple structure they can be summed.
In fact, by introducing the kernel
\be \label{eq:vec-kernel}
\cK(x,y) = \f{\l}{6 N} G_{ab}(x,y)G_{ba}(x,y)= \f{\l}{6} G(x,y)^2\;,
\ee
we find
\be \label{eq:vec-NLO}
 {\bf \G}^{(0)}_2[\phi,G] = \sum_{n\geq 1} \f{(-1)^{n+1}}{2n}  \Tr[\cK^n]= \f12 \Tr[\ln(\mathbf{1}+\cK)] \; .
\ee
Still at NLO, but for $\phi\neq 0$, we also have graphs like those of Fig.~\ref{fig:vectorNLO} but with exactly one solid propagator line being replaced by a background field insertion 
at each of its end vertices (e.g. the graph in Fig.~\ref{fig:graph}; see \cite{Berges:2004yj} for more details). Such graphs lead again to an action term which is  quadratic in $\phi$, thus not affecting the existence of the $\phi=0$ solution.

%-------------------------------------------------------------
\subsection{Large-$N$ expansion as the loop expansion of an auxiliary theory}
\label{sec:vec-auxiliary}
%-------------------------------------------------------------

Notice that \eqref{eq:vec-NLO} looks like the result we would obtain from a standard Gaussian integral with inverse covariance $\mathbf{1}+\cK$. 
It turns out that the large-$N$ expansion for the 2PI effective action of the vector model can indeed be cast as a loop expansion for an auxiliary bilocal theory, as we are now going to show.

Consider the partition function for the vector model, which corresponds to
\be
Z = e^{{\bf W}[0,0]} = e^{-{\bf \G}[\underline{\phi},\uG]} = \int [d\vph] \; e^{  - {\bf S}[\vph] } \;,
\ee
with the action \eqref{eq:vec-mod}. Next, insert in the functional integral the identity:
\be
\begin{split}
1 &= \int [d \tG] \d\left( N \tG(x,y) -  \vph_a(x) \vph_a(y)\right) \\
& = \int  [d \tG] [d\tilde\Sigma] \, e^{-\f12\int_{x,y}\tilde\Sigma(x,y)\left( N \tG(x,y) -  \sum_a \vph_a(x) \vph_a(y)\right)} \;,
\end{split}
\ee
exploit the delta function to write the interaction in terms of $\tG$, and then perform the integral over $\vph$:
\be\label{eq:Z-vec}
\begin{split}
Z & = \int  [d\vph][d \tG] [d\tilde\Sigma] \, e^{ - {\bf S}[\vph] -\f12\int_{x,y}\tilde\Sigma(x,y)\left( N \tG(x,y) -  \sum_a \vph_a(x) \vph_a(y) \right)}\\
&  = \int  [d \tG] [d\tilde\Sigma] \, e^{ - N\left\{ \f12 \Tr[(C^{-1}-\tilde\Sigma)\tG]  + \f12 \Tr[\ln (\tilde\Sigma) ]  +\f{\l}{4!} \int_x \tG(x,x)^2 \right\} }\\
&\equiv \int  [d \tG] [d\tilde\Sigma]  \, e^{ - N {\bf S}_{\rm eff}[\tG,\tilde\Sigma] } \;.
\end{split}
\ee

We have thus rewritten the original functional integral over $N$ (local) variables as an integral over just two (bilocal) variables, and all the dependence on $N$ is now explicit and factored in front of the total action.
Therefore, the $1/N$ expansion takes the standard form of a loop (i.e.\ saddle-point) expansion. 
We shift the fields to the saddle point value: $\tG(x,y) = \uG(x,y) + N^{-1/2} g(x,y)$, $\tilde\Sigma(x,y) = \uSigma(x,y) + N^{-1/2} \s(x,y)$.
Expanding to second order in $g$ and $\s$, we find:
\be
Z\simeq e^{ - N {\bf S}_{\rm eff}[\uG,\uSigma]}  \int  [d g] [d\s] \, e^{ - {\bf S}_{\rm eff}^{(2)}[\uG,\uSigma; g,\sigma]} 
 \;,
\ee
where the on-shell effective classical action coincides with the on-shell 2PI effective action at LO:
\be
{\bf S}_{\rm eff}[\uG,\uSigma]= \f12 \Tr[\ln \uG^{-1}] + \f12 \Tr[C^{-1}\uG] +  \f{\l}{4!}  \int_x \uG(x,x)^2\;.
\ee
We have also defined the quadratic part of the action:
\be
\begin{split}
{\bf S}_{\rm eff}^{(2)}[\uG,\uSigma; g,\sigma] =& - \f14 \int_{x_1,x_2,x_3,x_4} \s(x_1,x_2) \underline\cK_4(x_1,x_2;x_3,x_4) \s(x_3,x_4)\\
& - \f12 \int_{x_1,x_2} \s(x_1,x_2) g(x_1,x_2) +\f{\l}{4!} \int_x g(x,x)^2 \;,
\end{split}
\ee
where the kernel is: 
\be
\underline\cK_4(x_1,x_2;x_3,x_4) = \f12 ( \uG(x_1,x_3) \uG(x_2,x_4) +  \uG(x_1,x_4) \uG(x_2,x_3) ) \;.
\ee
Performing the Gaussian integrals we find:
\be
\begin{split}
Z &\simeq \f{e^{ - N {\bf S}_{\rm eff}[\uG,\uSigma]}}{ \left(  \det \left( \underline\cK_4  \right) \right)^{1/2}}  \int  [d g] \, e^{-\f14 \int_{x_1,x_2,x_3,x_4} g(x_1,x_2) \underline\cK_4^{-1}(x_1,x_2;x_3,x_4) g(x_3,x_4) -\f{\l}{4!} \int_x g(x,x)^2 }\\
&= e^{ - N {\bf S}_{\rm eff}[\uG] - \f12 \Tr[\ln(1+\underline{\cK})] } \;,
\end{split}
\ee
where the kernel $\underline{\cK}$ in the final result is exactly the one in \eqref{eq:vec-kernel} evaluated on shell.
Notice that due to the interaction being local rather than bilocal (compare with the SYK model in the next section), the 4-point kernel $\underline\cK_4$ reduces to the 2-point kernel $\underline{\cK}$.
We have thus recovered the LO and NLO of ${\bf \G}[\underline{\phi},\uG]$ by a standard saddle-point method.

A remark is in order. In ${\bf \G}[\underline{\phi},\uG]$ the on-shell fields
should be obtained from the full effective action. As we explained, $\underline\phi=0$ is valid to all orders in the symmetric phase, but the on-shell value $\uG$ receives 
corrections in $1/N$. Expanding ${\bf \G}[0,G]\simeq N {\bf \G}^{(1)}[0,G]+{\bf \G}^{(0)}[0,G]$, we find an expansion for the solution 
$\uG = \uG^{(0)}+N^{-1} \uG^{(-1)}$, and therefore, ${\bf \G}[\underline{\phi},\uG]=N{\bf \G}^{(1)}[0,\uG^{(0)}]+{\bf \G}^{(0)}[0,\uG^{(0)}]+O(N^{-1})$, because $\f{\d {\bf \G}^{(1)} }{\d G}[0,\uG^{(0)}]=0$ by construction.

\newpage
%-------------------------------------------------------------
\section{2PI effective action for the SYK model}
\label{sec:SYK}
%-------------------------------------------------------------

The SYK model is defined in terms of $N$ Majorana fermions in one dimension, with anti commutation relation $\{\psi_a,\psi_b\}=\d_{ab}$, and with action
\be
{\bf S}_{\rm SYK}[\psi] = \int dt \left(\f12  \psi_a \p_t \psi_a + \f{\im^{q/2}}{q!}  J_{a_1\ldots a_q} \psi_{a_1} \ldots \psi_{a_q} \right)\;.
\ee
Here, $J_{a_1\ldots a_q}$ is a random totally antisymmetric tensorial coupling, with Gaussian distribution
\be
P[J_{a_1\ldots a_q}] \propto \exp\left\{-\f{N^{q-1}(J_{a_1\ldots a_q})^2}{2(q-1)! J^2}  \right\} \;\;\;\;\;\text{(no sum)}\;.
\ee
We will denote with a bar the average over the disorder:
\be
\overline{A[J]} = \int \left(\prod_{a_1<a_2<\ldots <a_q}[d J_{a_1\ldots a_q}] P[J_{a_1\ldots a_q}]\right)  A[J]\;.
\ee
For example, we have
\be \label{eq:SYK-covariance}
\overline{J_{a_1\ldots a_q}J_{b_1\ldots b_q}} = \f{q!(q-1)!}{N^{q-1}} J^2 \,\Pi_{a_1\ldots a_q,b_1\ldots b_q}\;, %\;\;\text{(no sum)} \;,
\ee
where $\Pi_{a_1\ldots a_q,b_1\ldots b_q}$ is the projector on antisymmetric rank-$q$ tensors:
\be
\Pi_{a_1\ldots a_q,b_1\ldots b_q} = \frac{1}{q!} \sum_{\sigma\in \mathfrak{S}_q} \eps(\sigma) \prod_{i=1}^q \delta_{a_i b_{ \sigma(i)} } \;,
\ee
with $\mathfrak{S}_q$ the symmetric group on $q$ elements, and $\eps(\sigma)$ the sign of the permutation $\s$.

One deals with the randomness of the coupling by computing quenched averages of intensive quantities, such as the free energy or the entropy, which in general (e.g. for models with short-range interactions) are self-averaging, i.e.\ in the thermodynamic limit they converge with probability one to their average. In particular, the quenched free energy is
\be
-N\overline F = \overline{\ln Z} = \int \left(\prod_{a_1<a_2<\ldots <a_q}[d J_{a_1\ldots a_q}]  P[J_{a_1\ldots a_q}] \right) 
 \ln \int [d\psi] e^{-{\bf S}_{\rm SYK}[\psi]} \;.
\ee

The expansion in Feynman graphs is standard, with the only peculiarity that each vertex carries a tensor $J_{a_1\ldots a_q}$ with each index being associated to one half-edge.

In the same way as we defined a quenched free energy, we can define the quenched generating functionals of connected, 1PI, and 2PI diagrams, by constructing them in the usual
way for each realization of the disorder and taking the average over disorder at the end.
One should be careful with defining the generating functionals in such a way, because for example the averaging procedure does not in general commute with evaluating the effective action on shell. 
However, for the SYK model it can be shown by an analysis of the diagrams that commutativity holds at LO and NLO, a fact that here we will only show a posteriori by comparison to known 
results.\footnote{Note that in the standard way of obtaining LO and NLO results for the SYK model a replica diagonal ansatz is taken for the bilocal field, which is justified by the fact that for the SYK
model quenched and annealed averages coincide at LO and NLO \cite{Gurau:2017xhf,Kitaev:2017awl}. At NNLO, within the replica method one should take into account interactions between different replicas (i.e. off-diagonal 
fluctuations of the bilocal field), while in the 2PI formalism one should take into account diagrams that arise when the averaging is done after the on-shell evaluation.}

\begin{figure}[htb]
 \begin{center}
 \includegraphics[width=3cm]{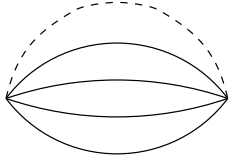}  
 \caption{The fundamental melon for $q=4$. The dashed line represents the Wick contraction associated to the quenched average.} \label{fig:fundMelon}
 \end{center}
 \end{figure}

We can therefore repeat all the construction of the 2PI effective action as above, with the novel feature that 2PI graphs contributing to ${\bf \G}_2$ now have to be averaged over disorder, and that the 
fermionic nature of the model brings in some minus factors.
We have
\be
{\bf \G}[\Psi,G] = {\bf S}_{\rm SYK}[\Psi]  - \f12 \Tr[\ln G^{-1}] - \f12 \Tr[G_0^{-1}G] +{\bf \G}_2[\Psi,G] \;.
\ee
In order to simplify the analysis of the large-$N$ limit we directly set $\Psi=0$, which is again justified by the absence of spontaneous symmetry breaking. By the same reason we could also fix $G_{ab}(x,y)=\d_{ab}G(x,y)$,
although in general it will be more transparent to keep the general expression.
After averaging over the disorder all the diagrams lead to different multiple traces of powers of $G_{ab}(t,t')$, and as before we should count each trace as contributing a factor $N$. We find in this way an expansion of 
the same type as \eqref{eq:vec-expansion}.
Remembering that in the large-$N$ limit the disorder average selects melons \cite{Kitaev}, we find that ${\bf \G}_2[0,G]$ 
at leading order in $1/N$ is given by the fundamental vacuum melon of Fig.~\ref{fig:fundMelon}, which is the only 2PI melon graph, with propagators given by $G$, i.e.:
\be \label{eq:SYK-LO}
\begin{split}
{\bf \G}_2^{(1)}[0,G] &=-\f{1}{2 q!} \overline{J_{a_1\ldots a_q}J_{b_1\ldots b_q}} \int_{t,t'} \prod_{c=1}^q G_{a_c b_c}(t,t') \\
&= -\f{J^2 }{2q N^{q-1}} \int_{t,t'} G_{aa}(t,t')^q = - \f{J^2 N}{2q} \int_{t,t'} G(t,t')^q  \;.
\end{split}
\ee
Notice that having chosen a Wick pairing of fermions to give the propagators (in $q!$ ways, thus canceling one of the $1/q!$ factors that come from the vertices), the average over disorder produces many different
types of contractions, due to the projector in \eqref{eq:SYK-covariance}, but in the second line we have taken the only contraction that contributes at LO.
The number of traces (and hence the power of $N$) is in general given by the number $c(\s)$ of cycles in the disjoint cycle decomposition of the permutation $\s$ appearing in the projector. Thus permutations that
can be obtained with a single transposition contribute to the NLO:
\be\label{eq:twisted-melon}
\begin{split}
&\left( -\f{1}{2 q!} \overline{J_{a_1\ldots a_q}J_{b_1\ldots b_q}} \int_{t,t'} \prod_{c=1}^q G_{a_c b_c}(t,t') \right)_{\rm NLO} 
\\ &\quad
 = -\f{J^2 }{2q N^{q-1}} \begin{pmatrix} q\\2\end{pmatrix} \int_{t,t'} G_{aa}(t,t')^{q-2} G_{bc}(t,t')G_{cb}(t,t')  = - \f{J^2 (q-1)}{4} \int_{t,t'} G(t,t')^q  \;.
\end{split}
\ee

The full 2PI effective action at leading order in $1/N$ is thus:
\be \label{eq:Gamma-SYK}
\f{1}{N} {\bf \G}[0,G] = - \f12 \Tr[\ln G^{-1}] - \f12 \Tr[\p_t G(t,t')] - \f{ J^2}{2q} \int_{t,t'} G(t,t')^q  \;,
\ee
which coincides with the bilocal action derived in \cite{Jevicki:2016bwu} by a change of variables within the replica method.
The equivalence with \cite{Jevicki:2016bwu} implies in particular that the equations of motion are the same and coincide with the SD equations:\footnote{Here one should remember that for Majorana fermions $G(t,t')=-G(t',t)$.}
\be \label{eq:SYK-SD}
G^{-1}(t,t') = \p_t \d(t,t') - J^2 G(t,t')^{q-1} \;,
\ee
and that the Schwarzian action controlling the conformal fluctuations can be derived in a similar fashion as in that paper.
We denote the solution of \eqref{eq:SYK-SD} as $\uG(t,t')$.

Recalling  Eq.~\eqref{eq:Wkk} and  ~\eqref{eq:truc}, the second derivative ${\bf \G}_{GG}[0,\underline{G}]$ is equal to  the inverse of  $\cF(t_1,t_2,t_3,t_4)$, i.e. of the full  4-point function minus the disconnected channel $(1\to 2,3\to 4)$. 
Interestingly, such channel is the leading-order (and uninteresting) term in the SYK 4-point function \cite{Kitaev,Maldacena:2016hyu}.\footnote{As a reminder, the 4-point function we are talking about is:
\be
\f{1}{N^2}\la \psi_m(t_1)\psi_m(t_2)\psi_n(t_3)\psi_n(t_4)\ra = G(t_{12})G(t_{34}) + \f1N \cF^{LO}(t_1,t_2,t_3,t_4) + \ldots \;.
\ee
The $G(t_{12})G(t_{34})$ part is precisely the channel missing when taking the derivatives as in \eqref{eq:Wkk}, and therefore, evaluating this derivative at LO will give us $\cF^{LO}$.
The latter was computed in \cite{Kitaev,Polchinski:2016xgd,Maldacena:2016hyu}.
}
Therefore, ${\bf \G}_{GG}[0,\underline{G}]$ captures precisely the inverse of the object of interest in SYK.
We can compute this from our LO effective action, and recover the corresponding result of the SYK 4-point function given by the sum of  the ladder diagrams, see \cite{Kitaev,Polchinski:2016xgd,Maldacena:2016hyu}.
Taking into account that:
\be  \label{eq:id-antisym}
\frac{\d G_{34}}{\d G_{12}} = \f12\left( \d(t_1-t_3)\d(t_2-t_4) - \d(t_1-t_4)\d(t_2-t_3) \right) \equiv  I_-(t_1,t_2 ; t_3, t_4) \;, 
\ee
with $I$ the orthogonal projector on antisymmetric functions, and denoting the on-shell four point kernel:
\be \label{eq:SYK-kernel}
\underline{\cK}(t_1,t_2;t_3,t_4) = - J^2 (q-1) \underline{G} (t_1,t_3) \underline{G} (t_2,t_4) \underline{G}(t_3,t_4)^{q-2} \;,
\ee 
 we get:
 %
%\be
\begin{align}
  {\bf \G}_{G_{34}} &= \frac{1}{2}\uG^{-1}(t_4,t_3) +\frac{1}{2} \p_t \d(t_3-t_4) - \frac{1}{2}J^2 [ \uG(t_3,t_4) ]^{q-1} \;, \\
 {\bf \G}_{G_{12}G_{34}}  & = - \frac{1}{4}\uG^{-1}(t_4,t_1)\uG^{-1}(t_2,t_3) + \frac{1}{4} \uG^{-1}(t_4,t_2)\uG^{-1}(t_1,t_3)  \crcr
    & \qquad  +  \frac{1}{4}[\delta(t_1-t_3) \delta(t_2-t_4) - \delta(t_1-t_4) \delta(t_2 - t_3)]  [- J^2 (q-1)   \underline{G} (t_3,t_4)^{q-2} ]\crcr
&=  -\f12  \int_{t,t'} \uG^{-1}(t_1,t)\uG^{-1}(t_2,t')  \bigg[  I_-(1 - \underline{\cK} )  \bigg](t,t';t_3,t_4) 
 \;.
\end{align}
%\ee
%
Inverting the last expression we find:
\be
\cF(t_1,t_2,t_3,t_4) =  \int_{t,t'}  \left(\f{1}{1-\underline{\cK}}\right)(t_1,t_2,t,t') (-\uG(t,t_3)\uG(t',t_4)+\uG(t,t_4)\uG(t',t_3))\;,
\ee
which is precisely the starting point of the computations in \cite{Kitaev,Polchinski:2016xgd,Maldacena:2016hyu}.

%-------------------------------------------------------------
\subsection{Next-to-leading order action}
%-------------------------------------------------------------

As in the vector model of the previous section, \eqref{eq:Gamma-SYK} will receive corrections at higher orders in $1/N$.
We want to show that the NLO correction can be interpreted as the result of performing the Gaussian integral over the fluctuations in the usual bilocal action expanded to quadratic order.
In order to do that, we need to understand which 2PI diagrams contribute at NLO, a question that has been addressed in detail in \cite{Bonzom:2017pqs} for the colored version of the model, which is a special 
case of the generalization of the SYK model introduced by Gross and Rosenhaus \cite{Gross:2016kjj} (see also \cite{Dartois:2017xoe} for a discussion of the same model at NLO).
It turns out that similar type of diagrams dominate also the standard SYK model, but have to be accompanied by the twisted melons \eqref{eq:twisted-melon}, which are absent in the colored case.

\begin{figure}[htb]
 \begin{center}
\includegraphics[width=6cm]{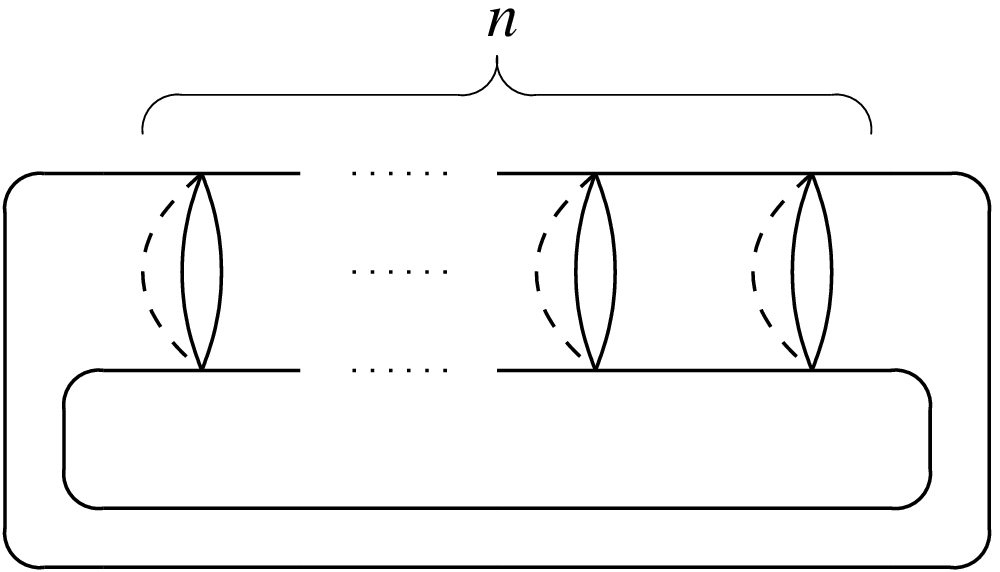}  
\hspace{1.5cm}
\includegraphics[width=6cm]{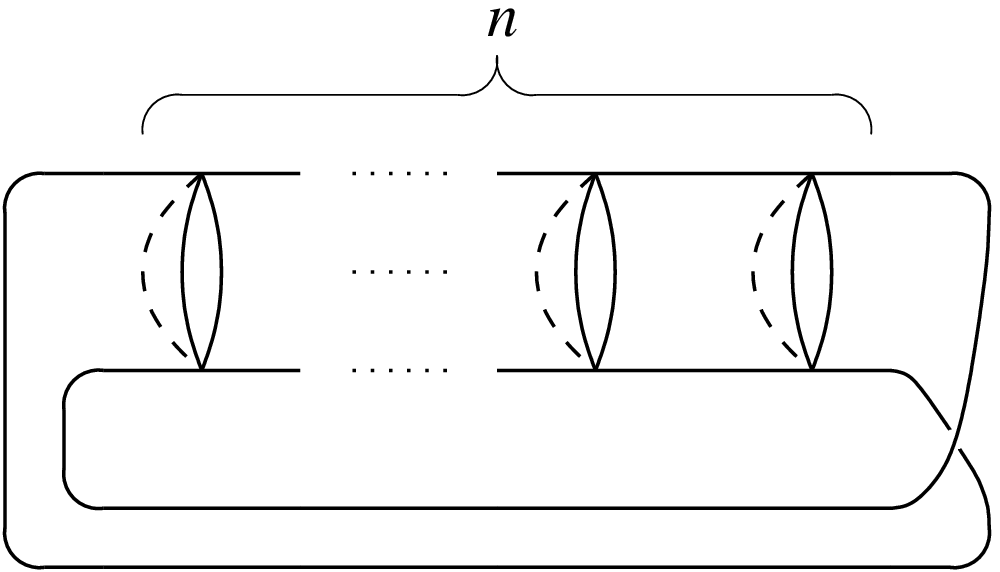}  
 \caption{NLO contributions with $n$ rungs, without (left) and with (right) twist.} \label{fig:SYK-NLO}
 \end{center}
 \end{figure}

The NLO 2PI vacuum graphs are thus given by all the periodic ladders with $n\geq 1$ rungs, with or without one twist of the rails, see Fig.~\ref{fig:SYK-NLO}.
One should notice that the case $n=1$ is quite special. First, the case $n=1$ without twist is 2-particle reducible if $q=4$, but since it evaluates to zero for any $q$ (because $G(t,t)=0$ due to the fermions' anti commutation), we can formally include it in the list. On the other hand, the case $n=1$ with twist corresponds again to a fundamental melon, thus one might think that it is LO rather than NLO. However, this corresponds precisely to the twisted melons in \eqref{eq:twisted-melon}, which therefore can be conveniently grouped with the ladders. 
Although such ladders form an infinite family of graphs, they can be summed in a similar way as to what we did for the vector case, i.e. by introducing a kernel for the insertion of a rung. One important difference is that now the kernel carries two vertices rather than one, which counts for different combinatorial factors and minus signs in the summation (notice that the kernel below has itself another minus sign, due to the fermionic nature of the theory). More explicitly, we have
\be \label{eq:SYK-NLO}
 {\bf \G}^{(0)}_2[0,G] = -\f12 \sum_{n\geq 0} \f{1}{n}  \Tr[\cK^n I_-]= \f12 \Tr[\ln(I_- -\cK I_- )] \; ,
\ee
where $I_-$ and $\cK$ are given in \eqref{eq:id-antisym} and \eqref{eq:SYK-kernel} (now off-shell), and to obtain the last equality we used the fact that $I_-=I_-^n$ because it is a projector, and $[\cK,I_-]=0$.
Evaluating ${\bf \G}^{(0)}_2[0,G] $ on the solution of the LO equations of motion, we find that this is the same result that one would obtain by integrating the quadratic fluctuations of the bilocal effective action of \cite{Maldacena:2016hyu,Kitaev:2017awl}.

In order to see that, we just have to repeat what we have done for the vector case, with the important difference that due to the disorder one has to use the replica method.
The quenched average in the SYK model can be performed exactly using the replica method, at the cost of introducing $n$ replicas of the system, and having to take the non-trivial limit $n\to 0$, which is needed in order to evaluate the
quenched free energy: $\overline{\ln Z} =  \lim_{n\to 0} \p_n \overline{Z^n}$.
One finds  \cite{Kitaev:2017awl}:
\be \label{eq:Z-replica}
\overline{Z^n} = \int \left(\prod_{\a\b} [d G^{\a\b}] [d\Sigma^{\a\b}]\right) e^{-N{\bf S}_{\rm eff}[G,\Sigma]} \;,
\ee
where:
\be \label{eq:Seff-replica}
{\bf S}_{\rm eff}[G,\Sigma] = -\f12  \widehat\Tr \ln(\p_t -\Sigma) + \f12\sum_{\a\b}\int_{t,t'} \left(\Sigma^{\a\b}(t,t') G^{\a\b}(t,t') - \f{J^2}{q}(G^{\a\b}(t,t'))^q\right) \;.
\ee
Notice that the disorder average led to an effective bare action which is bilocal even in the interaction term. 
Performing the saddle-point approximation with a replica diagonal ansatz $\underline G^{\a\b}=\underline G \d^{\a\b}$, which is valid up to NLO in $1/N$ \cite{Kitaev:2017awl}, one arrives at \cite{Maldacena:2016hyu}:
\be
\overline{\ln Z} =   N\left( \f12 \Tr[\ln \uG^{-1}] + \f12 \Tr[\p_t \uG(t,t')] + \f{ J^2}{2q} \int_{t,t'} \uG(t,t')^q \right)
- \f12 \Tr[\ln(I_- -\tilde\cK I_- )] \;,
\ee
with 
\be
\tilde\cK(t_1,t_2;t_3,t_4) = |\uG(t_1,t_2)|^{\f{q-2}{2}}\, \underline\cK(t_1,t_2;t_3,t_4)\, |\uG(t_1,t_2)|^{\f{2-q}{2}} \;.
\ee
Since $\f12 \Tr[\ln(I_- -\tilde\cK I_- )] = \f12 \Tr[\ln(I_- -\underline\cK I_- )] $, we recover our $ {\bf \G}[0,\uG]$ up to NLO, as claimed.

Note that the main difference between \eqref{eq:Seff-replica} with replica-diagonal ansatz and \eqref{eq:Z-vec} is that the SYK model the interaction part of ${\bf S}_{\rm eff}$ is bilocal while in the $O(N)$ model it is local. This is reflected in the fact that the associated fluctuation kernel is truly a 4-point kernel in the SYK case while it is a 2-point kernel in the $O(N)$ case.
From a graphical point of view the bilocality in the SYK model originates from the fact that the NLO graphs are ladders, while in the vector model they are chains of bubbles.

The replica diagonal ansatz used to derive the result above implies that $\overline{\ln Z}=\ln \overline{Z}$, i.e. that quenched and annealed averages coincide (see \cite{Gurau:2017xhf} for a combinatorial proof at LO).
Starting at NNLO \cite{Kitaev:2017awl}, the two averaging procedures start to differ, or in other words, the replica-symmetric ansatz becomes inaccurate.
From the point of view of the 2PI formalism, the complications at NNLO arise from the non-commutativity of averaging over disorder and going on shell, as we discussed before.

\newpage
%-------------------------------------------------------------
\section{2PI effective action for tensor field theories}
\label{sec:tensor2PI}
%-------------------------------------------------------------

A rank-$r$ tensor-valued real bosonic\footnote{In the fermionic case we will denote the field with $\psi$, and in dimensions $d>1$ one should remember also that its components are spinors.} 
field in $d$ space time dimensions is a function $\vph: \mathbb{R}^d \to \otimes_{i=1}^r V_i$, where $V_i$ is the vector space associated to the fundamental representation of a group $\cG_i$. In other words,
the tensor is postulated to be in the fundamental representation of a group $\cG=\prod_{i=1}^r \cG_i$.
We denote its components as $\vph_{a_1\ldots a_r}(x)$, with $x\in \mathbb{R}^d$ and $a_i=1\ldots N_i$, where $N_i={\rm dim}(V_i)$, hence the group acts by the transformation rule:
\be
\vph_{a_1\ldots a_r}(x) \to \left(\prod_{i=1}^r R^{(i)}_{a_ib_i}\right) \vph_{b_1\ldots b_r}(x)\;,
\ee
with the matrix $R^{(i)}$ belonging to the fundamental representation of the group $\cG_i$.\footnote{One could also consider tensors in an irreducible tensor representation of a single group, 
for example symmetric traceless or antisymmetric tensors for the group $\cG=O(N)$. Tensor models of this type (for rank $r=3$) have recently been proved to admit a large-$N$ expansion \cite{Benedetti:2017qxl}.}
We only consider tensor field theories defined by a classical action which is invariant under the action of $\cG$.

In the rest of the paper we will only study few specific models, yet we started this section with a very generic definition to emphasize that the construction of the 2PI effective action can be done in full generality.
In fact, it is straightforward to define the 2PI effective action for tensor-valued field theories in $d$ dimensions applying the construction that we reviewed in Sec.~\ref{sec:def-vector2PI}:
all the equations before Sec.~\ref{sec:vec-largeN} are in fact still valid, with the collective index now corresponding to an  $r$-uple of indices $(a_1\ldots a_r)$ plus the spacetime point.
For example, in rank 3 the bilocal field $G_{\bfa \bfb}$ corresponds to $G_{a_1 a_2 a_3 b_1 b_2 b_3}(x,y)$, and so on. 
The presence of several fields (as in the GW model defined below) is also straightforward to take into account: one simply needs to extend further the meaning of the vector label by including a field (or color) index $c=1\ldots q$. 
In this case, the discrete part of the collective-index $\bfa$ can be thought as a vector index with $\bfa=1\ldots M$, with $M=q\prod_{i=1}^r N_i$. The crucial property that characterizes a proper tensor model is the symmetry group: for a vector model the natural symmetry group would be $O(M)$, while for a tensor model this is broken  by the choice of interaction down to a smaller group with a natural tensorial interpretation (e.g. $O(M)$ is broken down to $O(N)^r$).

In the following we will consider only the cases $d=0$ and $d=1$, for two types of models: 
the Carrozza-Tanasa-Klebanov-Tarnopolsky (CTKT) model \cite{Carrozza:2015adg,Klebanov:2016xxf},
for winch $r=3$ and $\cG_i=O(N)$ for $i=1\ldots 3$, and the Gurau-Witten (GW) model \cite{color,Witten:2016iux},
in arbitrary rank $r=q-1$ and with  $\cG_i=O(N)$ for $i=1\ldots q(q-1)/2$.

%-------------------------------------------------------------
\subsection{The bosonic CTKT model in $d=0$}
%-------------------------------------------------------------

The CTKT model in zero dimensions is defined by the action:
\be \label{eq:CTKT-action}
{\bf S}_{\rm CTKT}[\vph] = \f12  \vph_{abc} \vph_{abc} + \f{\l}{4N^{3/2}} \vph_{a_1 a_2 a_3} \vph_{a_1 b_2 b_3}  \vph_{b_1 a_2 b_3} \vph_{b_1 b_2 a_3}  \;.
\ee
As standard, we refer to the location of an index as a color, e.g. the indices $a_1$ and $b_1$ in the action above are of color 1, and so on.

\begin{figure}[htb]
 \begin{center}
 \includegraphics[width=9cm]{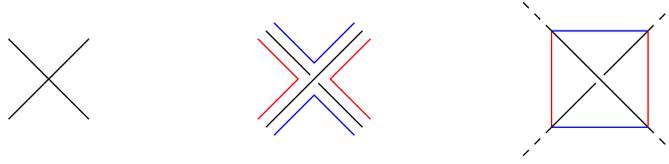}  
 \caption{The vertex of the CTKT model \eqref{eq:CTKT-action} in three different representation; from left to right: the Feynman representation, the stranded representation, and the ``tetrahedron'', or 
 edge-colored
 representation. In the last two the colors track the indices.} \label{fig:tensorVertex}
 \end{center}
 \end{figure}

The perturbative expansion can as usual be represented in a diagrammatic way. Due to the tensor structure, there are different possible representation, which we depict in Fig.~\ref{fig:tensorVertex}.

The corresponding  2PI effective action is constructed as as explained in Sec.~\ref{sec:def-vector2PI}, and in particular equation \eqref{eq:1loop+rest} is still valid, with ${\bf S}[\phi] = {\bf S}_{CTKT}[\phi] $,
\be
\begin{split}
& (G_0^{-1} )_{a_1 a_2 a_3 b_1 b_2 b_3} =   \d_{a_1b_1}\d_{a_2b_2}\d_{a_3b_3} \\
&\qquad + \f{\l}{N^{3/2}}  (\phi_{c_1 a_2 a_3} \phi_{c_1 b_2 b_3} \d_{a_1 b_1} + \phi_{a_1 c_2 a_3} \phi_{b_1 c_2 b_3} \d_{a_2 b_2}+ \phi_{a_1 a_2 c_3} \phi_{b_1 b_2 c_3} \d_{a_3 b_3}) \;,
\end{split}
\ee
and with ${\bf \G}_2[\phi,G]$ constructed from 2PI graphs with propagator $G_{a_1 a_2 a_3 b_1 b_2 b_3}$ and interaction
\be
{\bf S}_{\rm int}[\phi,\vph] = \f{\l}{N^{3/2}} \phi_{a_1 a_2 a_3} \vph_{a_1 b_2 b_3}  \vph_{b_1 a_2 b_3} \vph_{b_1 b_2 a_3}+ \f{\l}{4N^{3/2}} \vph_{a_1 a_2 a_3} \vph_{a_1 b_2 b_3}  \vph_{b_1 a_2 b_3} \vph_{b_1 b_2 a_3} \;.
\ee
As in the vector model, ${\bf \G}_2[\phi,G]$, and hence ${\bf \G}[\phi,G]$, will only contain even powers of $\phi$, and 
as a consequence, the equation of motion $\d{\bf \G}/\d\phi=0$ admits the solution $\phi=0$, which is the only solution giving an invariant 1-point function.
Thus we consider the case of zero background field, $\phi=0$, and study the large-$N$ expansion of ${\bf \G}_2[0,G]$.

\

\begin{figure}[htb]
 \begin{center}
 \includegraphics[width=8cm]{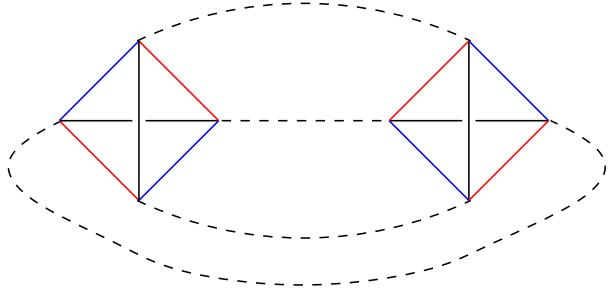}  
 \caption{The fundamental melon for the CTKT model in the tetrahedron representation. Dashed edges represent propagators.} \label{fig:fundMelon2}
 \end{center}
 \end{figure}

In order to do a large-$N$ expansion as in the vector case we need to identify quantities that scale like $N$.
In the vector case we saw that $\Tr[G^m]\sim N$ for any $m$. The easiest way to see such scaling is to assume that $G_{ab}\propto \d_{ab}$ which we know to be true for the on-shell 2-point function.
The analogue for the tensor case is to treat any ``trace'' over a given color as being of order $N$. Again the easiest way to see why it is so is to take $G_{a_1 a_2 a_3 b_1 b_2 b_3} \propto  \d_{a_1b_1}\d_{a_2b_2}\d_{a_3b_3}$,
which we know is going to be true on shell, due to the invariance of the theory.
The identification of the scaling with $N$ of the graphs contributing to the 2PI effective action is thus reduced to the well-studied problem of identifying the scaling with $N$ of tensor model graphs.
We can then borrow the results from \cite{Carrozza:2015adg}  and claim that :
\begin{itemize}
\item ${\bf \G}_2[0,G]$ can be expanded as:
\be \label{eq:KTCT-expansion}
 {\bf \G}_2[0,G] = \sum_{\omega\in \mathbb{N}/2} {\bf \G}^{ ( 3-\omega ) }_2[G]   \;,\qquad \text{with}\;\; {\bf \G}^{ (p) }_2[G] \sim N^p \;.
\ee
\item In the large-$N$ limit ${\bf \G}_2[0,G]$ is given by a single diagram, 
i.e. the fundamental vacuum melon (whose Feynman representation is the same as in Fig.~\ref{fig:fundMelon} with no dashed line, 
and whose tetrahedron representation is given in Fig.~\ref{fig:fundMelon2}), with propagators given by $G$: since the interaction is the known one, we know that melons dominate the large-$N$ limit,
and the fundamental melon is the only 2PI melon.\footnote{If we do not set $\phi=0$, we obtain in addition a 
term of the type $\phi^2G^3$, corresponding to a melon with two $\phi$ external legs and three internal propagators $G$. These terms should be taken into account when looking at fluctuations around the solution,
but at quadratic order the fluctuations of $\phi$ decouple from those of $G$ because $\underline\phi=0$ and the action is at least quadratic in $\phi$.}
\end{itemize}

Since the fundamental melon diagram comes with a combinatorial factor of 4, we obtain:
\be \label{eq:KTCT-LO-d0}
{\bf \G}_2^{(3)}[G] = -\f{\l^2}{8 N^3} G_{a_1 a_2 a_3 b_1 b_2 b_3} G_{a_1 a'_2 a'_3 b_1 b'_2 b'_3}  G_{a'_1 a_2 a'_3 b'_1 b_2 b'_3} G_{a'_1 a'_2 a_3 b'_1 b'_2 b_3} \;.
\ee

In the symmetric phase, the two point function is diagonal in the tensor indices:
\be
G_{a_1 a_2 a_3 b_1 b_2 b_3} = G\, \d_{a_1b_1}\d_{a_2b_2}\d_{a_3b_3}\;,
\ee
hence:
\be
{\bf \G}_2^{(3)}[G] = -\f18 \l^2 N^3  G^4  \; ,
\ee
and we obtain at leading order in $1/N$:
\be
\f{1}{N^3} {\bf \G}[0,G] =  \f12 \ln G^{-1} + \f12 G - \f18 \l^2  G^4 \; .
\ee
The LO equations of motion are simply:
\be
G^{-1} = 1 - \l^2  G^3 \;,
\ee
which we recognize as the SD equations at leading order in the $1/N$ expansion \cite{Carrozza:2015adg}.

\

\begin{figure}[htb]
 \begin{center}
 \includegraphics[width=3cm]{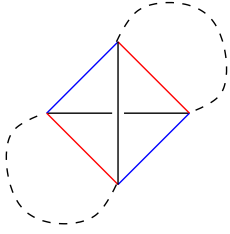}  
 \caption{The NLO core graph for the CTKT model in the tetrahedron representation.} \label{fig:CTKT-NLO}
 \end{center}
 \end{figure}

Following  \cite{Carrozza:2015adg}, one finds that at next-to-leading order the dominant graphs are generated by inserting melonic 2-point functions in the propagators of
the three core graphs obtained form the one depicted in Fig.~\ref{fig:CTKT-NLO} by  permutation of the colors. Since any insertion of a melonic 2-point function makes the graph 2-particle
reducible, we conclude that at NLO there is only a finite number of 2PI graphs, i.e. the three core graphs themselves.
They correspond to three contractions like:
\be
 \f{\l}{4 N^{3/2}}  G_{a_1 a_2 a_3 a_1 b_2 b_3} G_{a'_1 a_2 a_3 a'_1 b_2 b_3} \;.
\ee
With the diagonal ansatz for the two point function we obtain:
\be \label{eq:KTCT-NLO}
\f{1}{N^3} {\bf \G}[0,G] =  \f12 \ln G^{-1} + \f12 G - \f18 \l^2  G^4 + \f{3 \l}{4 N^{1/2}} G^2\;.
\ee
The equations of motion are now:
\be
G^{-1} = 1 - \l^2  G^3 +\f{3 \l}{N^{1/2}}  G\;.
\ee
Writing $ G = G^{(0)} + N^{-1/2} G^{(-1/2)}$ 
and expanding to order $N^{-1/2}$ we recover the SD equations at  NLO
of  \cite{Carrozza:2015adg}.\footnote{Up to a factor of 3 which was forgotten in \cite{Carrozza:2015adg}.}

\

We expect to find an infinite family of graphs at NNLO, but the analysis of the CTKT model at NNLO has never been done and it goes beyond the scope of the present paper. 
Since, on the contrary, the subleading structure of the GW model is much better understood, we will study the subleading corrections (up to NNNLO) in that model in Sec.~\ref{sec:GW}.

%-------------------------------------------------------------
\subsection{The fermionic CTKT model in $d=1$}
%-------------------------------------------------------------

Let us consider the $d=1$ fermionic CTKT model, which is in some ways the simplest tensor model with SYK-like large-$N$ limit \cite{Klebanov:2016xxf}.
Its classical action is:
\be
{\bf S}_{\rm CTKT}[\psi] = \int_t\left( \f12  \psi_{abc}(t) \p_t \psi_{abc}(t) + \f{\l}{4N^{3/2}} \psi_{a_1 a_2 a_3}(t) \psi_{a_1 b_2 b_3}(t)  \psi_{b_1 a_2 b_3}(t) \psi_{b_1 b_2 a_3}(t) \right) \;.
\ee
The selection of dominant graphs in the large-$N$ limit is not affected by the dimension of space time, hence the analysis of $d=0$ applies here without change. 
The Grassmann nature of the fields leads instead to some extra minus signs, just as in the SYK case.

We concentrate again on the symmetric phase $\underline{\Psi}=0$, which is the only possible one in $d=1$.
At LO in the $1/N$ expansion, ${\bf \G}_2[0,G]$ is given again by a single diagram,  the fundamental vacuum melon; with respect 
to \eqref{eq:KTCT-LO-d0} we only need to add the time dependence:
\be \label{eq:KTCT-LO-d1}
\begin{split}
{\bf \G}_2^{(3)}[G] &=  \f{-\l^2}{8 N^3} \int_{t,t'} G_{a_1 a_2 a_3 b_1 b_2 b_3}(t,t') G_{a_1 a'_2 a'_3 b_1 b'_2 b'_3}(t,t')  G_{a'_1 a_2 a'_3 b'_1 b_2 b'_3}(t,t') G_{a'_1 a'_2 a_3 b'_1 b'_2 b_3}(t,t')  \\
&= -\f18 \l^2 N^3 \int_{t,t'} G(t,t')^4  
\;,
\end{split}
\ee
where in the last equality we used a diagonal ansatz:
\be
G_{a_1 a_2 a_3 b_1 b_2 b_3} (t,t')= G(t,t') \d_{a_1b_1}\d_{a_2b_2}\d_{a_3b_3}\;,
\ee
which is valid on shell. By comparison with \eqref{eq:SYK-LO} it is obvious that we obtain the same behavior as in SYK, in particular the bilocal nature of the interaction.
In fact, including also the one-loop contribution:
\be \label{eq:KTCT-Gamma}
\f{1}{N^3} {\bf \G}[0,G] = - \f12 \Tr[\ln G^{-1}] - \f12 \Tr[\p_t G(t,t')] - \f18 \l^2  \int_{t,t'} G(t,t')^4 \;,
\ee
which has the same form as \eqref{eq:Gamma-SYK}.
If one were to not use a diagonal ansatz one would get:
\be \label{eq:KTCT-Gamma-nondiag}
{\bf \G}[0,G] = - \f12 \Tr[\ln G^{-1}_{a_1 a_2 a_3 b_1 b_2 b_3}] - \f12 \Tr[\p_t G_{a_1 a_2 a_3 b_1 b_2 b_3}(t,t')] + {\bf \G}_2^{(3)}[G] \;.
\ee
with ${\bf \G}_2^{(3)}[G]$ written as in the first line of \eqref{eq:KTCT-LO-d1}.

As pointed out in \cite{Choudhury:2017tax}, if in the infrared we discard the time-derivative term, the global $O(N)^3$ symmetry of \eqref{eq:KTCT-Gamma-nondiag} is promoted to a local symmetry.\footnote{Notice that this does not happen in the SYK model: in Eq.~\eqref{eq:SYK-LO} the trace $G_{aa}(t,t')$ identifies indices at different times, while in \eqref{eq:KTCT-LO-d1} indices are identified at equal times.}
The would-be gauge degrees of freedom associated to such local transformations are however proper degrees of freedom due to the explicit breaking provided by the time-derivative term, 
which we expect to endow them with an effective action controlling their dynamics. 
The idea is very similar to what happens with conformal symmetry: the action \eqref{eq:KTCT-Gamma} has precisely the same form as \eqref{eq:Gamma-SYK}, and as such it is also conformally invariant
(i.e. invariant under time reparametrizations) in the infrared/strong-coupling limit, i.e. when discarding the time-derivative term. The time derivative can then be viewed as a conformal breaking operator that generates 
an effective action for the conformal mode, which takes the form of a Schwarzian action \cite{Kitaev,Maldacena:2016hyu} (see  \cite{Jevicki:2016bwu,Jevicki:2016ito} for a derivation with an action with a single bilocal
field, as in our \eqref{eq:Gamma-SYK}, or \cite{Kitaev:2017awl} for more details on how to regularize the conformal breaking operator). 
Choudhury et al.~\cite{Choudhury:2017tax} have followed a similar route to obtain an effective action for the would-be gauge degrees of freedom, arriving at a non-linear sigma model type action, as one would expect on general grounds.
However, they postulated the action \eqref{eq:KTCT-Gamma-nondiag} as an effective classical action without any derivation, while we derived it here as a 2PI effective action. It is not clear at the moment whether a formulation analogous to the one in Sec.~\ref{sec:vec-auxiliary} exists for the KTCT model, but we can see two limitations to it: first, we expect such a formulation to be necessarily more complicated in the tensor case, because there are many more invariants, and the large-$N$ expansion cannot be interpreted as a loop expansion; second, as we saw in $d=0$, the NLO correction to the 2PI effective action of the KTCT model is given by a finite number of graphs and therefore it does not have the form of the result of a one-loop integral (compare \eqref{eq:KTCT-NLO} with \eqref{eq:vec-NLO} or \eqref{eq:SYK-NLO}), thus an hypothetical effective bilocal action would necessarily not factor the $N$-dependence as simply as in the vector case.
However, we can bypass such open question, and apply the same reasoning directly to the 2PI effective action. In order to see why, it is useful to recall that in Sec.~\ref{sec:SYK} we found that ${\bf \G}_{GG}$ gives the inverse 4-point function. 
The latter is then singular if ${\bf \G}_{GG}$ has zero eigenvalues, as it is the case if there is a gauge invariance which has not been gauge-fixed. In the present case we do not need a gauge fixing because there is an explicit breaking 
of the gauge invariance. The  would-be gauge modes give a non-zero contribution to the quadratic part of the action which can be obtained by evaluating the quadratic part of the breaking term in the gauge transformations around the stationary point.

In order to translate in formulas what we just said, we write:
\be \label{eq:Gamma_inv}
{\bf \G}_{\rm inv}[G] = - \f12 \Tr[\ln G^{-1}_{a_1 a_2 a_3 b_1 b_2 b_3}]+ {\bf \G}_2^{(3)}[G] \;,
\ee
\be \label{eq:Gamma_pert}
{\bf \G}_{\rm pert}[G] = - \f12 \Tr[\p_t G_{a_1 a_2 a_3 b_1 b_2 b_3}(t,t')]  \;.
\ee
The stationary point of the total action splits as (using boldface for a collective index only for the tensor indices, e.g. ${\bf a}=a_1a_2a_3$):
\be
\uG_{\bf a b} = \uG_0(t-t') \d_{\bf a b} +  \uG_1(t-t') \d_{\bf a b}\;,
\ee
where:
\be
\f{\d {\bf \G}_{\rm inv}}{\d G_{\bf a b}}[\uG_0] = 0\;,
\ee
\be \label{eq:G_1-pert}
\f{\d^2 {\bf \G}_{\rm inv}}{\d G_{\bf a b} \d G_{\bf cc}}[\uG_0]  \uG_1 + \f{\d {\bf \G}_{\rm pert}}{\d G_{\bf a b}}[\uG_0]= 0\;.
\ee
We emphasize that $ \uG_{\bf a b}$ is leading order in $1/N$: $\uG_1 $ is a perturbation in the strong coupling expansion, i.e. it arises by treating \eqref{eq:Gamma_pert} as a perturbation to  \eqref{eq:Gamma_inv}, but it is still leading order in the large $N$ limit.
Next, consider the transformation:
\be
G_{\bf a b} (t,t') \to G_{\bf a' b'} (t,t')  {\mathbb V}_{\bf aa'}(t){\mathbb V}_{\bf bb'}(t') \;,
\ee
where:
\be
{\mathbb V}_{\bf ac}(t) \equiv V^{(1)}_{a_1b_1}(t)V^{(2)}_{a_2b_2}(t)V^{(3)}_{a_3b_3}(t)  \simeq \d_{\bf ab}+{\mathbb H}_{\bf ab}(t) + \f12 {\mathbb H}_{\bf ac}(t){\mathbb H}_{\bf cb}(t) + \ldots \;,
\ee
\be
{\mathbb H}_{\bf ab}(t) 
= H_{a_1b_1}^{(1)}(t)\d_{a_2b_2}\d_{a_3b_3}+\d_{a_1b_1}H_{a_2b_2}^{(2)}(t)\d_{a_3b_3}+ \d_{a_1b_1}\d_{a_2b_2}H_{a_3b_3}^{(3)}(t)\;,
\ee
for $V_{ab}^{(i)}\in O(N)$ and $H_{ab}^{(i)}$ an antisymmetric matrix, for $i=1\ldots 3$.
Such transformation leaves ${\bf \G}_{\rm inv}[G] $ invariant, but not ${\bf \G}_{\rm pert}[G] $.
Using the invariance of the former, and the linearity in $G$ of the latter, it can be easily shown (expanding at first order in $\uG_1$ the left-hand-side and using \eqref{eq:G_1-pert}) that:
\be \label{eq:trickGoldstone}
\f{\d^2( {\bf \G}_{\rm inv}+ {\bf \G}_{\rm pert})}{\d G_{\bf a b} \d G_{\bf cd}}[\uG] g_{\bf ab}  g_{\bf cd} 
= \f{\d^2 {\bf \G}_{\rm pert}[\uG_0 {\mathbb V}_{\bf ac}{\mathbb V}_{\bf bc}]}{\d {\mathbb H}_{\bf a b} \d {\mathbb H}_{\bf cd} } {\Big|_{{\mathbb H}=0}}{\mathbb H}_{\bf a b}  {\mathbb H}_{\bf cd} \;,
\ee
where:
\be
g_{\bf ab} = \underline{G}(t,t') ({\mathbb H}_{\bf ab}(t)-{\mathbb H}_{\bf ab}(t'))\;.
\ee
Rewriting the quadratic part of $ \underline{G}(t-t')  {\mathbb V}_{\bf ab}(t){\mathbb V}_{\bf ab}(t')$ as:
\be
\begin{split}
& \f12 \underline{G}(t-t') ({\mathbb H}_{\bf ac}(t){\mathbb H}_{\bf ca}(t) + {\mathbb H}_{\bf ac}(t'){\mathbb H}_{\bf ca}(t')-2{\mathbb H}_{\bf ab}(t){\mathbb H}_{\bf ba}(t')) \\
& \quad \simeq \f12 \underline{G}(t-t') \left(\p_t {\mathbb H}_{\bf ac}(t) \p_t{\mathbb H}_{\bf ca}(t) (t-t')^2 + O((t-t')^3)\right)
\;,
\end{split}
\ee
we obtain:
\be
\f{\d^2 {\bf \G}_{\rm pert}[\uG_0 {\mathbb V}_{\bf ac}{\mathbb V}_{\bf bc}]}{\d {\mathbb H}_{\bf a b} \d {\mathbb H}_{\bf cd} } {\Big|_{{\mathbb H}=0}}{\mathbb H}_{\bf a b}  {\mathbb H}_{\bf cd} 
=  -\f{\a}{2} \int_t \p_t {\mathbb H}_{\bf ac}(t) \p_t{\mathbb H}_{\bf ca}(t) \;,
\ee
where
\be
\a = \int_\t \uG_0(\t) \t^2 \s(\t)\;,
\ee
with $\s(\t)$ a suitable regularization of $\d'(\t)$. This is precisely the same coefficient that appears in front of the Schwarzian action, as derived in \cite{Kitaev:2017awl}, and the action coincides with the one derived in \cite{Choudhury:2017tax}.

%-------------------------------------------------------------
\subsection{The fermionic GW model in $d=1$}
\label{sec:GW}
%-------------------------------------------------------------

\begin{figure}[htb]
 \begin{center}
 \includegraphics[width=7cm]{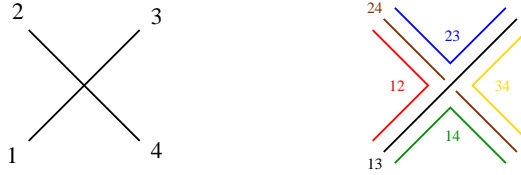}  
 \caption{The vertex of the GW model \eqref{eq:GW-action} with $q=4$ in the Feynman (left) and the stranded (right) representations. } \label{fig:colorVertex}
 \end{center}
 \end{figure}

The (real) GW model in one dimension is defined by the action:
\be \label{eq:GW-action}
{\bf S}_{\rm GW}[\psi] = \f12 \sum_{c=1}^q \int_{t,t'}  \psi^{(c)}_{\bf a_c}(t) C^{-1}(t,t') \psi^{(c)}_{\bf a_c}(t') + \f{\im^{q/2} \l }{N^{(q-1)(q-2)/4}} \int_t 
\prod_{c=1}^q\psi^{(c)}_{\bf a_c}(t)  \prod_{c_1<c_2} \d_{a_{c_1 c_2} a_{c_2 c_1}}  \;,
\ee
where ${\bf a_c}=(a_{c c_1} | c_1 \in \{1,\ldots,q\}\backslash \{c\})$ and $C^{-1}(t,t')=\p_{t} \d(t-t')$. The vertex is represented in Fig.~\ref{fig:colorVertex} for the case $q=4$. 
The model is symmetric under the global group $O(N)^{q(q-1)/2}$, where an independent $O(N)$ element acts on each pair $(a_{c_1 c_2},a_{c_2c_1})$.

We introduce a bilocal source for each color and obtain the 2PI effective action:
\be
{\bf \G}[\Psi^{(c)},G^{(c)}] = {\bf S}_{\rm GW}[\Psi^{(c)}]  + 
\f{1}{2} \sum_{c=1}^q \Tr[\ln (G^{(c)} )] - \f{ 1 }{2} \sum_{c=1}^q\Tr[(G_0^{(c)})^{-1}G^{(c)}] +{\bf \G}_2[\Psi^{(c)},G^{(c)}] \;.
\ee
From now on we consider the symmetric phase $\underline{ \Psi}^{(c)}=0$. 
The leading order 2PI graph is again the fundamental melon, thus ${\bf \G}_2[0,G^{(c)}]$ has a large-$N$ expansion which starts at order $N^{q-1}$:
\be
\begin{split}
{\bf \G}_2^{(q-1)}[0,G^{(c)}] &= - \f{\l^2}{2 N^{(q-1)(q-2)/2}} \int_{t,t'} \prod_{c=1}^q G^{(c)}_{\bf a_c b_c}(t,t') \prod_{c_1<c_2} \d_{a_{c_1 c_2} a_{c_2 c_1}} \d_{b_{c_1 c_2} b_{c_2 c_1}} \\
&= - \f{\l^2 N^{q-1}}{2} \int_{t,t'} G(t,t')^q \;,
\end{split}
\ee
where in the last step we restricted to the color symmetric and diagonal ansatz $G^{(c)}_{\bf a_c b_c}(t,t') = G(t,t') \prod_{c'\neq c} \d_{a_{cc'}b_{cc'}}$. With such ansatz we recover precisely the SYK result \eqref{eq:Gamma-SYK}, with $J$ replaced by $\l$ and $N$ by $qN^{q-1}$.

\

As $ - {\bf \G}_2$ is the sum over 2PI vacuum graphs, the subleading corrections  begin at order $N^2$ for any $q$ \cite{Bonzom:2017pqs}, and the diagrams contributing to $ {\bf \G}_2^{(2)} $, $ {\bf \G}_2^{(1)}$ and  $ {\bf \G}_2^{(0)}$ are all the ring graphs 
 consisting in 4-point ladder diagrams closing onto themselves, similar to the ones depicted in Fig.~\ref{fig:SYK-NLO}, but in which we need to distinguish the various possible sequences of colors along the rails  \cite{Bonzom:2017pqs}. 
In order to compute their contributions to the 2PI effective action, we recall that the Gaussian expectation with covariance $G$ of real fermions is:
\be
 \Braket{ \psi(t_{1} ) \dots \psi(t_{ 2n } ) } = \sum_{\pi} \epsilon(\pi) \prod_{(k,l)\in \pi} G(t_{k},t_{l}) \;,
\ee
where $\pi$ are the pairings of $2n$ elements, $(k,l)$ with $k<l$ and $\epsilon(\pi)$ is the signature of the pairing. 
We are interested in the perturbative expansion at order $2n$:
\be
\frac{\lambda^{2n} \im^{nq}}{(2n)!}  \Braket{ \prod_{i=1}^{2n}      \psi^1(t_i) \dots \psi^q(t_i)   }_{\rm ring}\;,
\ee
where the subscript signals that we only select the contractions that reproduce ring graphs.

A ring graph is built by first pairing the $2n$ vertices into $n$ pairs where the vertices in a pair are connected by $q-2$ edges.
We denote the colors of the external edges of a dipole $c_1$ and $c_2$.
For each pair we get a factor:
\begin{equation}
\begin{split}
 & (-1)^{q/2 + q(q-1)/2} \psi^{c_1}( t ) \psi^{c_2}(t) \bigg[ G( t,t')  \bigg]^{q-2} \psi^{c_2}(t' )  \psi^{c_1}(t' ) \crcr
& \qquad \qquad  = \psi^{c_1}( t )  \psi^{c_1}(t' )  (-1)  \bigg[ G( t,t')  \bigg]^{q-2} \psi^{c_2}(t' ) \psi^{c_2}(t)  \;.
\end{split}
\end{equation}
We now glue the pairs together to form ring graphs. This identifies the right external colors on a pair with the left external colors on the next pair. 
The field at $t_i$ can connect with either $t_{i+1}$ or $t'_{i+1}$ and we obtain schematically:
\begin{equation}
\begin{split}
&   \frac{\lambda^{2n}}{(2n)!} \frac{(2n)!}{2^n n!} (n-1)! 2^{n-1}\sum_{\{ c_i\} }^{c_i\neq c_{i-1}} \bigg( \prod_{i=1}^n G(t_{i-1},t_i) G(t'_{i-1}, t'_i) (-1)  \bigg[ G( t_i,t'_i)  \bigg]^{q-2} \bigg) \crcr
& \qquad \qquad \qquad  \bigg[ \delta(t_n-t_0) \delta(t'_{n} - t'_0) - \delta(t_n - t'_0) \delta(t'_n - t_0) \bigg] \;. 
\end{split}
\end{equation}
We now reinstate the tensor indices. 
We denote:
\begin{align}
 & \hat {\mathbb K}^{(c_1 c_2)}_{ {\bf a_{c_1} a'_{c_1} } ; { {\bf b_{c_2} b'_{c_2} } }} (t_a,t_{a'} ; t_{b}, t_{b'}) = \crcr
 & \qquad =  (-1) G^{(c)}_{{\bf a_{c_1} b_{c_1} } }(t_a,t_b)  G^{(c)}_{{\bf a'_{c_1} b'_{c_1} }}(t_{a'},t_{b'}) 
\left(  \prod_{c \neq c_1,c_2} G^{ ( c ) }_{ {\bf  b_{c} b'_{c} } } (t_b,t_{b'})  \right) \left( \prod_{c < c'} \delta_{b_{cc'} b_{c'c}} \delta_{b'_{cc'} b'_{c'c}} \right) \;,\crcr 
 & \mathbb{I}^{=}_{  _{ {\bf a_{c} a'_{c} } ; { {\bf b_{c} b'_{c} } }}   }(t_a,t_{a'} ; t_b , t_{b'})  
 = \delta_{ {\bf a_{c} b_{c}  }} \delta_{ {\bf a'_{c} b'_{c}  }} \delta(t_a - t_b) \delta(t_{a'} - t_{b'})  \;, \crcr
&  \mathbb{I}^{\times}_{  _{ {\bf a_{c} a'_{c} } ; { {\bf b_{c} b'_{c} } }}   }(t_a,t_{a'} ; t_b, t_{b'})   = \delta_{ {\bf a_{c} b'_{c}  }} \delta_{ {\bf a'_{c} b_{c}  }} 
\delta(t_a - t_{b'}) \delta(t_{a'} - t_b ) \;,
\end{align}
where repeated indices are summed. Later on we will take the color symmetric diagonal ansatz for the two point function. We denote:
\begin{align}
 \hat \cK (t_a,t_{a'} ; t_b, t_{b'}) & = (-1) G(t_a,t_b) G(t_{a'},t_{b'}) [G(t_b,t_{b'})]^{q-2}  \;, \crcr 
 I^{=}(t_a,t_{a'} ; t_b, t_{b'}) & = \delta(t_a - t_b) \delta(t_{a'} - t_{b'})  \;, \crcr
 I^{\times}(t_a,t_{a'} ; t_b , t_{b'}) & = \delta(t_a - t_{b'}) \delta(t_{a'} - t_b) \;,
\end{align}
and $\hat \cK I^{\times} = I^{\times} \hat \cK$.
As a function of the sequence of horizontal colors $c_i$,  as well as the last contraction, we get the following contributions to the 2PI effective action:
\begin{align}
  {\bf \G}_2^{(2)} & = - \sum_{n\ge 2}  \frac{\lambda^{2n}}{2n}   N^{  - n (q-1) (q-2) / 2 }   \;\sum_{(c_1\dots c_n)\in U_n}^{c_{n+1} = c_1} \Tr\bigg[  \left( \prod_{i=1}^n \hat {\mathbb K}^{(c_ic_{i+1})} \right)  {\mathbb I}^{=}\bigg]  \;, 
 \crcr
  {\bf \G}_2^{(1)} & = - \sum_{n\ge 2}   \frac{\lambda^{2n}}{2n}  N^{  - n (q-1) (q-2) / 2 }   \;\sum_{(c_1\dots c_n)\in U_n}^{c_{n+1} = c_1} \Tr\bigg[  \left( \prod_{i=1}^n \hat {\mathbb K}^{(c_ic_{i+1})} \right) ( - {\mathbb I}^{\times} )\bigg] \;,
 \\ \nn
  {\bf \G}_2^{(0)} & = - \sum_{n\ge 2}  \frac{\lambda^{2n}}{2n} N^{  - n (q-1) (q-2) / 2 }   \;\sum_{(c_1\dots c_n)\in B_n}^{c_{n+1} = c_1} \Tr\bigg[  \left( \prod_{i=1}^n \hat {\mathbb K}^{(c_ic_{i+1})} \right) ( {\mathbb I}^{=} - {\mathbb I}^{\times} )\bigg] \;,
\end{align}
where $U_n$ is the set of alternating (or \emph{unbroken}) words $(c_1 c_2 \dots c_1 c_2)$ of length $n$ with $c_1<c_2$ over the colors, and 
$B_n$ is the set of non alternating (or \emph{broken}) words $ (c_1 \dots c_3 \dots c_2)\;, \;\; c_i\neq c_{i+1}$ of length $n$ with $c_1<c_2$ over the colors.
Restricting to the color symmetric diagonal ansatz we get:
\begin{align}
 {\bf \G}_2^{(2)} & = - N^2 \sum_{n\ge 2} \frac{\lambda^{2n}}{2n}  \; | U_n |  \Tr\bigg[ \hat \cK^n I^{=}\bigg]   \;,
 \crcr
 {\bf \G}_2^{(1) } & = - N \sum_{n\ge 2}   \frac{\lambda^{2n}}{2n} \; | U_n | \Tr\bigg[ \hat \cK^n  (- I^{\times} ) \bigg]  \;,
 \\ \nn
 {\bf \G}_2^{(0) } & = - \sum_{n\ge 2}  \frac{\lambda^{2n}}{2n} \; | B_n | \Tr\bigg[ \hat \cK^n (I^{=} - I^{\times} )\bigg]   \;.
\end{align}
In all these cases $ - N^{-r} \lambda \partial_{\lambda}{\bf \G}^{(r)}_2$ is a generating function of nonempty words with weight $\lambda^2 \hat \cK$ per letter. 
Ignoring for an instant the fact that $\hat \cK$ is an operator and denoting in superscript the two external letters of the word we have:
\begin{itemize}
 \item{\it Unbroken words.}
  The generating functions of nonempty, unbroken words are simple geometric series:
  \begin{equation}
   U^{c_1c_1} =  \frac{ (q-1) \lambda^6 \hat \cK^3}{1 - \lambda^4 \hat \cK^2} \;,\qquad U^{c_1c_2} = \frac{\lambda^4 \hat \cK^2 }{1 - \lambda^4\hat \cK^2}  = \lambda\partial_{\lambda} \bigg[ -  \frac{1}{4}\ln(1 - \lambda^4 \hat \cK^2) \bigg] \;,   
  \end{equation}
 \item { \it Arbitrary words.} The generating function of nonempty, arbitrary words with equal external letters is:
 \[
  A^{c_1c_1} = \lambda^2 \hat \cK \frac{ (q-1) \lambda^2 \hat \cK  }{ 1 - (q-2) \lambda^2 \hat \cK }   (\lambda^2 \hat \cK + A^{c_1c_1}) = 
  \frac{ (q-1) \lambda^6 \hat \cK^3  }{ 1 - (q-2) \lambda^2 \hat \cK -  (q-1) \lambda^4 \hat \cK^2    }
  \;,
 \]
because an arbitrary, non empty word with equal external letters $c_1$ is: a letter $c_1$ followed by a nonempty word which does not reuse the letter $c_1$,
followed by either exactly a letter $c_1$ or a nonempty word with external letters $c_1 c_1$. 
The generating function of nonempty, arbitrary words with different external letters is:
  \begin{equation}
  \begin{split}
   A^{c_1c_2} & =  \lambda^2 \hat \cK \bigg[ \frac{  1 }{1 - (q-2) \lambda^2 \hat \cK  } \bigg]  (\lambda^2 \hat \cK  + A^{c_2c_2}) =  \frac{ \lambda^4 \hat \cK^2 }{ 1 - (q-2) \lambda^2 \hat \cK   -  (q-1) \lambda^4 \hat \cK^2  } \crcr
      & \qquad  = \lambda \partial_{\lambda} \bigg[ - \frac{1}{2q(q-1)}  \ln[1 - (q-1) \lambda^2 \hat \cK] - \frac{1}{2q} \ln (1+ \lambda^2 \hat \cK )   \bigg] \;,      
  \end{split}
  \end{equation}
  as such a word is a letter $c_1$ followed by a possibly empty word which does not use the letter $c_2$, followed by 
  either a letter $c_2$ or a nonempty word with external letters $c_2c_2$.
 \item {\it Broken words.}  The generating function of nonempty, broken words with different external letters is: 
\begin{equation}
 \begin{split}
  & B^{c_1c_2} = A^{c_1c_2} - U^{c_1c_2} = \crcr
  & \qquad =  \lambda \partial_{\lambda} \bigg[ - \frac{1}{2q(q-1)}  \ln[1 - (q-1) \lambda^2 \hat \cK] - \frac{1}{2q} \ln(1+\lambda^2 \hat \cK)  + \frac{1}{4} \ln(1 - \lambda^4 \hat \cK^2)  \bigg] \;.
 \end{split}
\end{equation}
\end{itemize}
Recalling now that $\hat \cK$ is an operator and that the projector on antisymmetric functions is $I_- =(I^{=} - I^{\times} )/ 2$ 
we get with the color-symmetric diagonal ansatz:
\be \label{eq:fullGamma-GW}
{\bf \G}[0,G]  =  N^{q-1} \frac{q}{2} \Tr [\ln(G)] - N^{q-1}\frac{q}{2} \Tr[\partial_t G] + {\bf \G}_2[G] \;,
\ee
where:
\be
{\bf \G}_2[G] = {\bf \G}_2^{(q-1)}[G] + {\bf \G}_2^{(2)}[G] + {\bf \G}_2^{(1)}[G] + {\bf \G}_2^{(0)}[G]\;,
\ee
and:
 \begin{align}\label{eq:explicit}
   {\bf \G}_2^{(q-1)}[G] &  = N^{q-1}  \left( - \frac{\lambda^2}{2}  \right) \int_{t,t'} G(t,t')^{q}  \;,\crcr
   {\bf \G}_2^{(2)}[G] & =N^2 \frac{1}{4} \binom{q}{2} \Tr\bigg[  I^{=}  \ln\bigg( 1 - \lambda^4 \hat \cK^2 \bigg) \bigg] \;,
   \crcr
    {\bf \G}_2^{(1)}[G] & = N \frac{1}{4} \binom{q}{2}\Tr\bigg[ ( - I^{\times} ) \ln\bigg( 1 - \lambda^4 \hat \cK^2 \bigg)   \bigg] \;,
    \\ \nn
     {\bf \G}_2^{(0)}[G] & = \frac{1}{2}  \Tr\bigg[  I_-  \ln\bigg( 1 - (q-1)\lambda^2 \hat \cK \bigg) \bigg] \;, \crcr \nn
     & \qquad +  \frac{q-1}{2}  \Tr\bigg[ I_-  \ln\bigg( 1 + \lambda^2 \hat \cK \bigg)  \bigg] - \frac{1}{2} \binom{q}{2}\Tr\bigg[ I_- \ln\bigg( 1 - \lambda^4 \hat \cK^2 \bigg)   \bigg] \;.
 \end{align}

A first use of Eq.~\eqref{eq:explicit} is to determine the two point function at subleading order in $1/N$. 
For instance, truncating the equation $\partial_G {\bf \G} =0$ at leading and next to leading order for the bosonic GW model in $d=0$ by we obtain:
\begin{equation}
  0= - G^{-1} + 1 - \lambda^2 G^{q-1} - \frac{1}{N^{q-3}} \binom{q}{2} \frac{ \lambda^4 G^{2q-1}}{ 1 -\lambda^4 G^{2q} } \;,
\end{equation}
and substituting $G = \underline{G}^{(0)} + N^{-q+3 } \underline{G}^{( -q+3)}$, with $ \underline{G}^{(0)} = 1 + \lambda^2 ( \underline{G}^{(0)})^q$, we get:
\begin{equation}
\underline{G}^{( -q+3)} = \frac{1}{N^{q-3} } \binom{q}{2} \frac{ \lambda^4 (\underline{G}^{(0)} )^{2q }}{ [ 1 -\lambda^4 ( \underline{G}^{(0)})^{2q} ] [ 1 - q \lambda^2 (\underline{G}^{(0)})^{q-1} ] } \;,
\end{equation}
reproducing the result of \cite{GurSch}.

Going back to $d=1$ we observe that $\partial_{G}{\bf \G}^{(q-1)}|_{G = \underline{G}^{(0)}} =0$, 
hence   ${\bf \G}[0, \underline{G}^{(0)} + N^{-q+3 } \underline{G}^{( -q+3)} ] = {\bf \G}[0, \underline{G}^{(0)}]  $ up to terms of order 
$N^{q-1} (N^{-q+3 } \underline{G}^{( -q+3)})^2  \sim N^{5-q} (\underline{G}^{( -q+3)})^2$, that is (except for $q=4$ which is special) subleading with respect to all the terms in Eq.~\eqref{eq:explicit}.
Thus, up to order $N^0$, the free energy of the GW model with $q\geq 6$ is (introducing also the projector on symmetric functions $I_+ =(I^{=} + I^{\times} )/ 2$):
\begin{equation} \label{eq:GW-freeEn}
 \begin{split}
 -\ln Z  = & N^{q-1} \frac{q}{2} \Tr [\ln(\uG^{(0)})] - N^{q-1}\frac{q}{2} \Tr[\partial_t \uG^{(0)}] - N^{q-1} \frac{\lambda^2}{2} \int_{t,t'} \underline{G}^{(0)}(t,t')^{q}  \\
&   + \bigg[ \frac{N(N-1)}{2} \binom{q}{2} \bigg] \frac{1}{2} \Tr\bigg[ \ln\bigg( 1 - \lambda^4 [ \underline{\hat \cK}^{(0)}]^2 I_+\bigg) \bigg] \\
  &  + \bigg[ \bigg(\frac{N(N-1)}{2} + (N-1)\bigg) \binom{q}{2} \bigg] \frac{1}{2} \Tr \bigg[ \ln \bigg( 1 - \lambda^4 [ \underline{\hat \cK}^{(0)}]^2 I_-\bigg) \bigg]  \\
  &+ (q-1) \frac{1}{2}   \Tr\bigg[ \ln\bigg( 1 + \lambda^2 [ \underline{\hat \cK}^{(0)}] I_- \bigg)   \bigg]  +  \frac{1}{2}  \Tr\bigg[ \ln\bigg( 1 - (q-1)\lambda^2 [\underline{\hat \cK}^{(0)}] I_- \bigg)   \bigg] \;,
 \end{split}
\end{equation}
where the four point kernel $\underline{\hat \cK}^{(0)}$ is evaluated on $ \underline{G}^{(0)} $, the on-shell leading-order two point function, and where we have rearranged the subleading terms in order to eliminate $I^{=}$ and $I^{\times}$ in favor of $I_\pm$. For $q=4$ the terms of order $N$ and order 1 receive corrections from $\uG^{(-1)}$.
All the subleading correction have the form of traces of a logarithm, hence each of them can be interpreted as resulting from the integration of freely fluctuating bilocal fields. 
Furthermore, the factor $\frac{N(N-1)}{2} \binom{q}{2}$ in the second line is very suggestive of the number of antisymmetric matrices on color $ij$, $i\neq j$, while the factor $\big(\frac{N(N-1)}{2} + (N-1)\big) \binom{q}{2}$ 
in the third line is suggestive of the number of symmetric traceless matrices on the same colors. Such interpretation is in fact correct, as we will now show.

It turns out that the final result \eqref{eq:GW-freeEn} can be interpreted as a one-loop approximation for a bilocal effective action of the same form as the 2PI effective action at LO:
\be \label{eq:Seff-GW}
\begin{split} 
{\bf S}_{\rm eff}[G] = & \f{1}{2} \sum_{c=1}^q \Tr[\ln (G^{(c)} )] - \f{ 1 }{2} \sum_{c=1}^q\Tr[(G_0^{(c)})^{-1}G^{(c)}] \\
& - \f{\l^2}{2 N^{(q-1)(q-2)/2}} \int_{t,t'} \prod_{c=1}^q G^{(c)}_{\bf a_c b_c}(t,t') \prod_{c_1<c_2} \d_{a_{c_1 c_2} a_{c_2 c_1}} \d_{b_{c_1 c_2} b_{c_2 c_1}} \;.
\end{split}
\ee
In order to see that, we split the bilocal field as on-shell background plus fluctuations,
\be
G^{(c)}_{\bf a_c b_c}(t,t') = \uG^{(0)}(t,t') \d_{\bf a_c b_c} + g^{(c)}_{\bf a_c b_c}(t,t')\;, 
\ee
and expand the action to second order in the fluctuations $g^{(c)}_{\bf a_c b_c}(t,t')$.
We obtain a quadratic action of the form (see Appendix \ref{app:decomp} for notation):
\be \label{eq:gg-action}
\la {\bf g} | \mathbb{B}( \mathbb{I} -    \lambda^2 \mathbb{K}  )  | {\bf g} \ra \;,
\ee
where $\mathbb{B}$ is a $\l$-independent $q\times q$ block matrix with $(\uG^{(0)})^{-1}(\uG^{(0)})^{-1}$ on its diagonal and zero otherwise. The latter leads to a $\f12 \Tr\ln\mathbb{B}$ term in the free energy that should be canceled by the measure, as for zero coupling ${\bf \G}_2[G]$ should vanish. Notice that in the case of the $O(N)$ and SYK models  we obtained the correct measure thanks to the Lagrange multiplier $\tilde\Sigma$; we could introduce a similar field here by analogy, but since we are not deriving ${\bf S}_{\rm eff}$ directly from the path integral it seems more natural to just fix the normalization by the zero-coupling condition.
Therefore, we can replace $\mathbb{B}=1$ in \eqref{eq:gg-action}.

\begin{figure}[htb]
 \begin{center}
 \includegraphics[width=3.5cm]{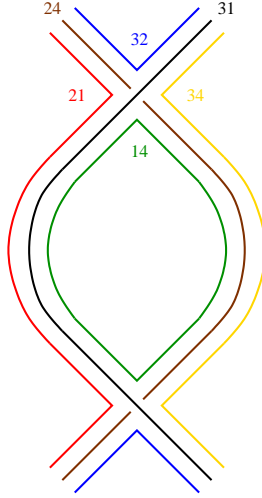}  
 \caption{The kernel $\cK^{(c_1 c_2)}$ for $q=4$ and $(c_1 c_2)=(2 3)$. When ``gluing'' to its right a fluctuation $g^{(3)}_{a_{31}a_{32}a_{34} b_{31}b_{32}b_{34} }$, with for example the $a$ indices on top and the $b$ indices at the bottom, the indices of color 31 and 34 are traced, while the index of color 32 is transmitted. } \label{fig:dipole}
 \end{center}
 \end{figure}

The important point to notice is that the operator  $\mathbb{K} $ in \eqref{eq:gg-action} is built out of kernels $\cK^{(c_1 c_2)}$ that when acting on $g^{(c_1)}_{\bf a_c b_c}(t,t')$ or $g^{(c_2)}_{\bf a_c b_c}(t,t')$ take their trace 
with respect to all the indices of color different from $c_1 c_2$ (see Fig.~\ref{fig:dipole}). 
%%%%%%%%%%%
 %%%%%%%%%%%
Therefore, it is useful to decompose (see Appendix \ref{app:decomp} for details):
\be \label{eq:g-decomp}
g^{(c)}_{\bf a_c b_c}(t,t') = g^{(c)}(t,t') \prod_{i\neq c} \d_{a_{ci} b_{ci}}+ \sum_{i\neq c} g^{(ci)}_{a_{ci} b_{ci}}(t,t') \prod_{j\neq i,c} \d_{a_{cj} b_{cj}} + \hat{g}^{(c)}_{\bf a_c b_c}(t,t') \;,
\ee
 where $g^{(ci)}_{a_{ci} a_{ci}}(t,t')=0$, for any $i$, and  $\hat{g}^{(c)}_{\bf a_c b_c}(t,t') \prod_{j\neq i} \d_{a_{cj} b_{cj}}=0$, for any $i\neq c$.
 
One can then further decompose $g^{(ci)}_{a_{ci} b_{ci}}(t,t')$ in symmetric traceless and antisymmetric parts with respect to the matrix indices (since $g^{(ci)}_{a_{ci} b_{ci}}(t,t')= - g^{(ci)}_{b_{ci} a_{ci}}(t',t)$, 
the symmetry properties with respect to $t$ and $t'$ are opposite to those of the indices).

The Hessian has a block-diagonal form corresponding to the decomposition \eqref{eq:g-decomp}.
The block corresponding to the scalar modes $g^{(c)}(t,t')$ is a $q\times q$ matrix with the identity operator $I_-$ on the diagonal entries, 
and $-\l^2\underline{\hat \cK}^{(0)} I_-$ on the off-diagonal ones. Such a matrix has one eigenvalue $(1-(q-1)\l^2\underline{\hat \cK}^{(0)})I_-$,
and $(q-1)$ eigenvalues $(1+\l^2\underline{\hat \cK}^{(0)})I_-$, thus upon integration of such modes we obtain the last line in \eqref{eq:GW-freeEn}.
The blocks corresponding to the matrix modes $g^{(c_1 c_2)}_{a_{ci} b_{ci}}(t,t')$, are instead $2\times 2$ matrices for each fixed pair $c_1c_2$, with the identity operator $\Im \equiv \cS I_- + \cA I_+$ (the operators $\cS$ and $\cA$ are the projectors on symmetric traceless and antisymmetric matrices, respectively) on the diagonal entries, and $-\l^2\underline{\hat \cK}^{(0)} \Im$ on the off-diagonal ones. Integration over such modes thus produces the second and third line of  \eqref{eq:GW-freeEn}.
Lastly, the block corresponding to $\hat{g}^{(c)}_{\bf a_c b_c}(t,t')$ is just the identity, hence it does not lead to any subleading correction to the free energy.

%-------------------------------------------------------------
\section{Summary and outlook}
\label{sec:concl}
%-------------------------------------------------------------

We have introduced and discussed the 2PI effective action for the SYK model and for tensor field theories.
The main lessons we drew from that are:
\begin{itemize}

\item For the SYK model, the 2PI effective action easily reproduces all the the results of the bilocal action formalism \cite{Maldacena:2016hyu,Jevicki:2016bwu,Kitaev:2017awl}, without using the replica method, at least up to the same order in $1/N$ at which the replica symmetric ansatz works for the latter.

\item For tensor analogues of the SYK model, the 2PI formalism offers so far the only way to obtain an effective action for collective fields, and it allows to obtain the same type of results as in the SYK model.

\item For the CTKT model, the 2PI effective action provides a solid starting point for the argument of \cite{Choudhury:2017tax}, showing the existence of soft modes associated to the $O(N)^3$ quasi-gauge invariance in the strong coupling limit. A similar argument can be repeated straightforwardly for the GW model with the symmetry group being replaced by  $O(N)^{q(q-1)/2}$.

\item For the GW model, the $1/N$ expansion of the 2PI effective action can be pushed up to NNNLO, and for all three subleading orders we find traces of logarithms, which have a natural interpretation as the result of Gaussian integrals over bilocal fields. Somewhat surprisingly, such Gaussian integrals correspond precisely to the one-loop approximation for a bilocal effective action of the same form as the leading-order 2PI effective action.

\end{itemize}
 
 We think that the 2PI formalism is particularly promising for the exploration of subleading effects in $1/N$ in tensor field theories. Hopefully this can lead to a better understanding of the underlying degrees of freedom and their possible holographic interpretation. It would also be interesting to carry out a NNLO analysis for the CTKT model to uncover similar trace log terms.
 
 \newpage
 
 \appendix
 
 \section{Orthogonal decomposition of the fluctuations}\label{app:decomp}
 
To simplify notation let us suppress the time variables. We organize the fluctuations $g^{(c)}_{{\bf a_c} {\bf b_c}}$ in a column vector
with $q$ entries, each of which is a $N^{q-1} \times N^{q-1}$ matrix:
\be
{\bf g} = \begin{pmatrix}
     g^{(1)}_{{\bf a_1} {\bf b_1}} \\     g^{(2)}_{{\bf a_2} {\bf b_2}} \\ \vdots     \\  g^{(q)}_{{\bf a_q} {\bf b_q}}  
 \end{pmatrix} \;, \qquad  \Braket{ {\bf h} | {\bf g} } = \sum_{c=1}^q \Tr[ (h^{(c)})^T  g^{(c)} ] \;,
\ee
where $T$ denotes transposition.
We denote $\delta_{  {\bf a_{c} b_{c} }  } \equiv \prod_{c'\neq c}  \delta_{a_{cc'} b_{cc'}}$ and $\delta^{c  -  c c_1}_{  {\bf a_{c} b_{c} }  } \equiv \prod_{c'\neq c,c_1}  \delta_{a_{cc'} b_{cc'}}$.
The identity operator in this vector space writes:
\be
\mathbb{I} =\begin{pmatrix}
             I^{(1)}  & 0 &  0 \\ 0 & \ddots & 0 \\  0 & 0 &  I^{(q)} 
            \end{pmatrix} \;, 
\qquad  I^{(c)}_{ {\bf a_{c} b_{c} } ; {\bf m_{c} n_{c} }    } =  \delta_{  {\bf a_{c}   m_{c} }  } \delta_{  {\bf b_{c}  n_{c}  }  }  \;.
\ee
The 4-point kernel  is the operator:
\be
\begin{split}
 \lambda^2 \mathbb{K} & = \lambda^2 \begin{pmatrix}
                  0            & K^{(1 \, 2)} & \dots  & K^{(1  \, q ) } \\ 
                  K^{(2 \, 1)} & 0           & \dots  & K^{ (2 \; q) } \\
                  \vdots       &             &        &            \\
                  K^{(q \, 1)} & K^{(q \, 2)}&  \dots  & 0 
               \end{pmatrix} \; ,  
  \crcr 
  K^{(c_1c_2)}_{   {\bf a_{c_1} b_{c_1} }; {\bf m_{c_2} n_{c_2} }  } & =\frac{ {\hat K} }{  N^{q-2}  } \;  \delta^{c_1 - c_1 c_2}_{  {\bf a_{c_1} b_{c_1} }  }   \delta_{ a_{c_1c_2} m_{c_2c_1}   } \delta_{b_{c_1c_2}   n_{c_2c_1}  }  
    \delta^{c_2 - c_2c_1}_{  {\bf m_{c_2} n_{c_2} }  }   \; ,
\end{split}
\ee
and the Hessian of \eqref{eq:Seff-GW} is proportional to the linear operator $\mathbb{I} -    \lambda^2 \mathbb{K} $ on this vector space. 
Notice that when $K^{(c_1c_2)}$ acts on $ g^{(c_2)}_{  {\bf m_{c_2} n_{c_2} }  }$ it traces it on $q-2$ indices, i.e.\ those of color different from $c_1c_2$ (see again Fig.~\ref{fig:dipole}), thus showing the way to its partial diagonalization.
We introduce the following operators:
\be 
P^{ ( c_1 c_2 ) }_{ {\bf a_{c_1} b_{c_1} } ; {\bf m_{c_2} n_{c_2} }   }   = \frac{1}{   N^{q-1}  } \; \delta_{  \bf a_{c_1} b_{c_1}  } \delta_{  \bf m_{c_2} n_{c_2}  }  \;,
\ee 
%Tracing over $q-2$ indices of $ g^{(c)}_{  {\bf a_{c} b_{c} }  }$,  one obtains $(q-1)$ different $N\times N$ matrices (one for each choice of color on which we do not trace). 
%The projector on the traceless part of such matrices is denoted:
%
\be
 T^{ ( c_1 c_2 ) }_{   \bf a_{c_1} b_{c_1}  ;  \bf m_{c_2} n_{c_2}  } = \frac{1}{  N^{q-2}  } \;  \delta^{c_1- c_1c_2}_{ \bf a_{c_1} b_{c_1}  }  
   \bigg( \delta_{ a_{c_1c_2}  m_{c_2c_1} } \delta_{  b_{c_1c_2} n_{c_2c_1} }  -\frac{1}{N} \delta_{  a_{c_1c_2}  b_{c_1c_2} }  \delta_{ m_{c_2c_1}n_{c_2c_1} }   \bigg)   \delta^{c_2- c_2c_1}_{ \bf m_{c_2} n_{c_2}  } \; ,
\ee
and:
\be
R^{(c-c c_1)}_{   \bf a_{c} b_{c}  ;  \bf m_{c} n_{c}  } = \frac{1}{  N^{q-2}  } \;  \delta^{c- c c_1}_{ \bf a_{c} b_{c}  }  
   \bigg( \delta_{ a_{cc_1}  m_{cc_1} } \delta_{  b_{cc_1} n_{cc_1} }  -\frac{1}{N} \delta_{  a_{cc_1}  b_{cc_1} }  \delta_{ m_{cc_1}n_{cc_1} }   \bigg)   \delta^{c- cc_1}_{ \bf m_{c} n_{c}  } \;.
\ee
In words, when acting on a fluctuations $g^{(c_2)}$, $P^{ ( c_1 c_2 ) }$ traces all the indices and replaces them with an identity on color $c_1$; $T^{ ( c_1 c_2 ) }$ does the same but spares the shared color $( c_1 c_2 )$, on which it projects on the traceless part; lastly, $R^{(c_2-c_2 c_1)}$ is similar to $T^{ ( c_1 c_2 ) }$, but it does not change the color of the traced indices.
They satisfy (no sum over $c$): 
%$P^{ ( c_1 c ) }P^{ ( c c_2 ) }=P^{ ( c_1 c_2 ) }$,  $T^{ ( c_1 c ) }T^{ ( c c_2 ) }=R^{ (c_2- c_2 c ) }\d_{c_1c_2}$, $R^{(c-c c_1)}R^{(c-c c_2)}=R^{ (c- c c_2 ) }\d_{c_1c_2}$, and  $P^{ ( c_1 c ) }T^{ ( c c_2 ) }= T^{ ( c_1 c ) }P^{ ( c c_2 ) }=0$.
%
\begin{align}
P^{ ( c_1 c ) }P^{ ( c c_2 ) }&=P^{ ( c_1 c_2 ) }\;,  & P^{ ( c_1 c ) }T^{ ( c c_2 ) }&= T^{ ( c_1 c ) }P^{ ( c c_2 ) }=0 \;,\crcr
T^{ ( c_1 c ) }T^{ ( c c_2 ) } &=R^{ (c_2- c_2 c ) }\d_{c_1c_2}\;, & T^{ ( c_1 c ) }R^{(c-c c_2)} &= T^{ ( c_1 c) } \d_{c_1c_2}\\ \nn
 R^{(c-c c_1)}R^{(c-c c_2)}&=R^{ (c- c c_2 ) }\d_{c_1c_2}\;, & P^{ ( c_1 c ) }R^{ (c-c c_2) }&= R^{(c-c c_1) }P^{ ( c c_2 ) }=0\;.
\end{align}
In the vector space spanned by ${\bf g}$  the 4-point kernel splits as the sum of two operators $ \mathbb{K} = {\hat K} (  \mathbb{T}  + \mathbb{P}  )  $,
with:
\be
 \mathbb{P}   = \begin{pmatrix}
                  0            & P^{(1 \, 2)} & \dots  & P^{(1  \, q ) } \\ 
                  P^{(2 \, 1)} & 0           & \dots  & P^{ (2 \; q) } \\
                  \vdots       &             &        &            \\
                  P^{(q \, 1)} & P^{(q \, 2)}&  \dots  & 0 
               \end{pmatrix} \; ,   \qquad  
                         \mathbb{T}  = \begin{pmatrix}
                  0            & T^{(1 \, 2)} & \dots  & T^{(1  \, q ) } \\ 
                  T^{(2 \, 1)} & 0           & \dots  & T^{ (2 \; q) } \\
                  \vdots       &             &        &            \\
                  T^{(q \, 1)} & T^{(q \, 2)}&  \dots  & 0 
               \end{pmatrix} \; .
\ee 

Introducing also the projectors:
\be
 \mathbb{Q}   = \begin{pmatrix}
             P^{(1 \, 1)} & 0 &  0 \\ 0 & \ddots & 0 \\  0 & 0 &   P^{(q \, q)} 
            \end{pmatrix} \; ,   
\qquad  
\mathbb{R}  = \mathbb{T}^2 = \begin{pmatrix}
            \sum_{c\neq 1} R^{(1-1 c)} & 0 &  0 \\ 0 & \ddots & 0 \\  0 & 0 &   \sum_{c\neq q} R^{(q-q c)} 
            \end{pmatrix} \; ,
\ee 
the identity can be decomposed in orthogonal components as $\mathbb{I} = \hat{\mathbb{I}} +  \mathbb{Q}  +  \mathbb{R} $, where $\hat{\mathbb{I}}=\mathbb{I} -  \mathbb{Q}  -  \mathbb{R} $.
Using such a decomposition of the identity we can write:
\be
{\bf g} =  \mathbb{Q}{\bf g}+  \mathbb{ R }{\bf g} + \hat{\mathbb{I}}{\bf g} \; ,
\ee
which in components is:
\be
g^{(c)}_{\bf a_c b_c} = g^{(c)}  \delta^{c}_{  \bf a_{c} b_{c}  }  + \sum_{c_1\neq c} g^{(cc_1)}_{a_{cc_1} b_{cc_1}} \delta^{c-cc_1}_{  \bf a_{c} b_{c}  }  + \hat{g}^{(c)}_{\bf a_c b_c} \;,
\ee
with: 
\be
\begin{split}
  g^{(c)} & = \frac{1}{N^{q-1}}  \delta^{c}_{  \bf m_{c} n_{c}  }  g^{(c)}_{  \bf m_{c} n_{c}  }   \;, \crcr
   g^{(cc_1)}_{a_{cc_1} b_{cc_1}} & = \frac{1}{N^{q-2}}
   \bigg( \delta_{ a_{cc_1}  m_{cc_1} } \delta_{  b_{cc_1} n_{cc_1} }  -\frac{1}{N} \delta_{  a_{cc_1}  b_{cc_1} }  \delta_{ m_{cc_1}n_{cc_1} }   \bigg)   \delta^{c- cc_1}_{ \bf m_{c} n_{c}  }   g^{(c)}_{  \bf m_{c} n_{c}  } \;, \crcr
   \delta^{c-cc_1}_{\bf a_c b_c}  \hat{g}^{(c)}_{\bf a_c b_c}  & = 0\; ,
\end{split}
\ee
which is the decomposition introduced in \eqref{eq:g-decomp}. 

The quadratic action for the fluctuations thus writes:
\be \label{eq:diagHess}
\la {\bf g} | ( \mathbb{I} -    \lambda^2 \mathbb{K}  )  | {\bf g} \ra = 
\la \mathbb{Q}{\bf g} | ( \mathbb{Q} -    \lambda^2 {\hat K} \mathbb{P}  )  | \mathbb{Q}{\bf g} \ra 
+ \la \mathbb{R}{\bf g} | ( \mathbb{R} -    \lambda^2 {\hat K} \mathbb{T}  )  | \mathbb{R}{\bf g} \ra 
+ \la \hat{\mathbb{I}}{\bf g} |  \hat{\mathbb{I}}{\bf g} \ra \;.
\ee
Furthermore, we can decompose:
\be \label{eq:RgRg}
\la \mathbb{R}{\bf g} | ( \mathbb{R} -    \lambda^2 {\hat K} \mathbb{T}  )  | \mathbb{R}{\bf g} \ra =
N^{q-2} \sum_{c_1<c_2} \left(g^{(c_1 c_2)} \; g^{(c_2 c_1)}\right)   \begin{pmatrix} 1 & - \l^2 {\hat K} \\  - \l^2 {\hat K}  & 1 \end{pmatrix}
\begin{pmatrix} g^{(c_1 c_2)} \\  g^{(c_2 c_1)}\end{pmatrix} \;.
\ee
Lastly, we notice that given that $g^{(c c')}_{a  b }(t,t')=-g^{(c c')}_{ b a }(t',t)$ , we can rewrite (omitting the subscript $c c'$ on the indices):
\be \label{eq:gg-Im}
\begin{split}
g^{(c c')}_{a  b }(t,t')  g^{(c c')}_{a  b } (t,t')
= \int_{s,s'}g^{(c c')}_{a  b }(t,t') \Im_{a  b ; m  n } (t,t';s,s') g^{(c c')}_{m  n } (s,s')\;,
\end{split}
\ee
where:
\be
\Im_{a  b ; m  n } (t,t';s,s') =  \cS_{a  b ; m  n } I_-(t,t';s,s') + \cA_{a  b ; m  n } I_+(t,t';s,s')\;,
\ee
and:
\begin{align}
\cS_{a  b ; m  n } &=\frac{1}{2} (\delta_{a m} \delta_{b n}+\delta_{a n} \delta_{b m}) \;,\\
\cA_{a  b ; m  n } &=\frac{1}{2} (\delta_{a m} \delta_{b n}-\delta_{a n} \delta_{b m})\;,\\
I_\pm(t,t';s,s') &= \f12\left( \d(t-s)\d(t'-s') \pm \d(t-s')\d(t'-s) \right) \;.
\end{align}
Together, \eqref{eq:diagHess}, \eqref{eq:RgRg}, and \eqref{eq:gg-Im} realize the block-diagonalization described in the text, thus leading to the trace log of \eqref{eq:GW-freeEn} upon integration over the fluctuations.\footnote{Plus some constant factors (in particular logarithmic terms in $N$ coming for example from the $N^{q-2}$ factor in \eqref{eq:RgRg}), which can be absorbed in the measure.}

%%%%%%%%%%%%%%%%%

%----- Bibliography ----------------------
%\bibliographystyle{JHEP-3}
%\bibliography{refs}
%---------------------------------------------

\providecommand{\href}[2]{#2}\begingroup\raggedright\endgroup

%------------------------------------------------------------------------------
\end{document}